\documentclass[journal=jacsat,manuscript=article]{achemso}

\usepackage[version=3]{mhchem} 
\usepackage{xcolor}

\newcommand*{\red}{\textcolor{black}}
\author{Kamal Choudhary}
\email{kamal.choudhary@nist.gov}
\affiliation[National Institute of Standards and Technology]
{Material Measurement Laboratory, National Institute of Standards and Technology, Gaithersburg, 20899, MD, USA.}
\alsoaffiliation[Theiss Research]
{Theiss Research, La Jolla, 92037, CA, USA.}
\alsoaffiliation[DeepMaterials LLC]
{DeepMaterials LLC, Maryland, 20906, MD, USA.}

 \author{Mathew L. Kelley }
\affiliation[National Institute of Standards and Technology]
{Physical Measurement Laboratory, National Institute of Standards and Technology, Gaithersburg, 20899, MD, USA.}
\alsoaffiliation[Theiss Research]
{Theiss Research, La Jolla, 92037, CA, USA.}

\title{ChemNLP: A Natural Language Processing based Library for Materials Chemistry Text Data}

\abbreviations{IR,NMR,UV}
\keywords{American Chemical Society, \LaTeX}

\begin{document}

\begin{tocentry}
\includegraphics[width=8cm,height=4.5cm]{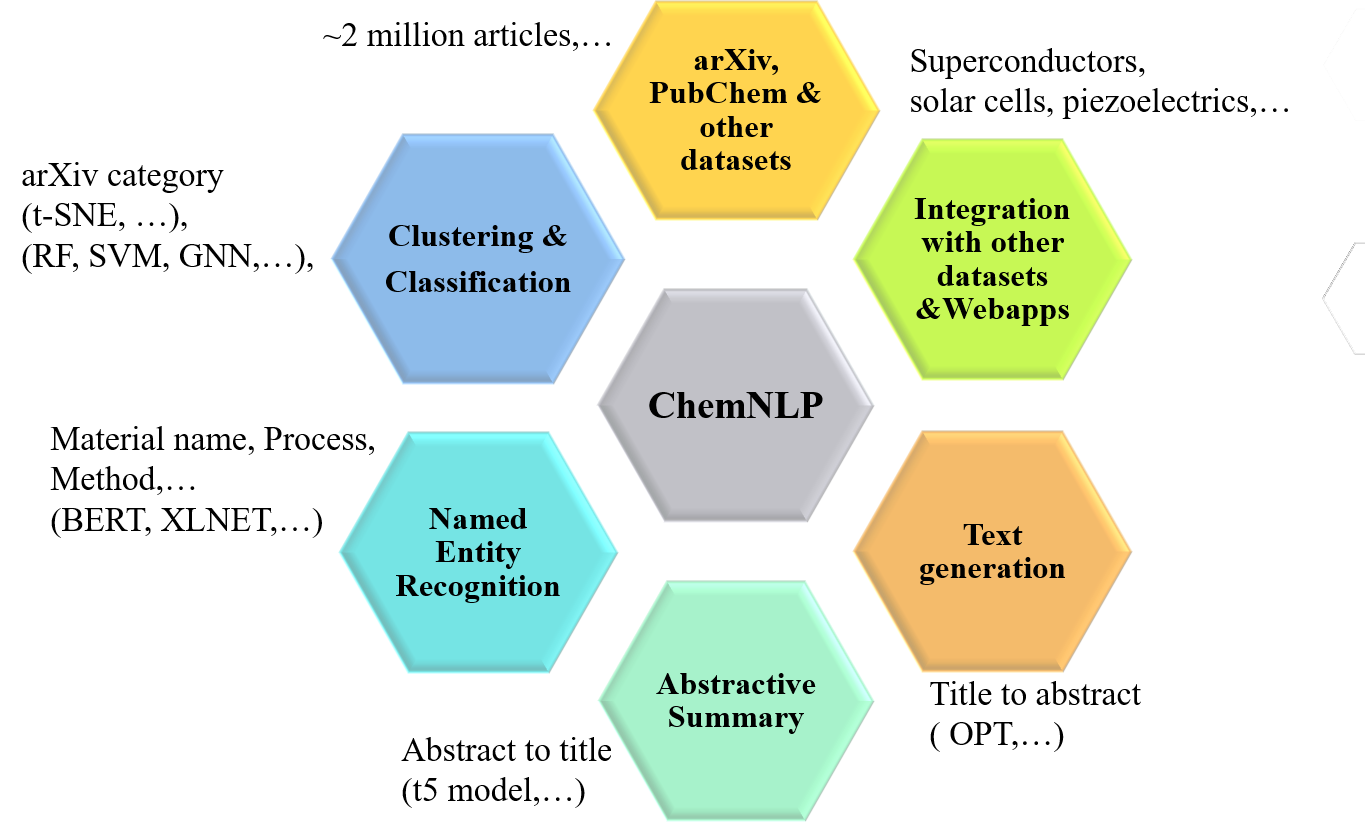}
\end{tocentry}

\begin{abstract}

In this work, we present the ChemNLP library that can be used for 1) curating open access datasets for materials and chemistry literature, developing and comparing traditional machine learning, transformers and graph neural network models for 2)  classifying and clustering texts, 3) named entity recognition for large-scale text-mining, 4) abstractive summarization for generating titles of articles from abstracts, 5) text generation for suggesting abstracts from titles, 6) integration with density functional theory dataset for identifying potential candidate materials such as superconductors, and 7) web-interface development for text and reference query. We primarily use the publicly available arXiv and Pubchem datasets but the tools can be used for other datasets as well. Moreover, as new models are developed, they can be easily integrated in the library. ChemNLP is available at the websites: \url{https://github.com/usnistgov/chemnlp} and \url{https://jarvis.nist.gov/jarvischemnlp/}.

\end{abstract}

\section{Introduction}

The number of scholarly articles available on the web is estimated to be more than 100 million \cite{jinha2010article,khabsa2014number}. It is an overwhelming task to perform a specific scientific query and extract meaningful information from such a large corpus. Natural language processing (NLP) is a subfield of artificial intelligence and linguistics to make computers understand the statements or words written in human languages and perform useful tasks \cite{chowdhary2020natural,mikolov2013distributed}. NLP can be used on scholarly articles for several applications such as text summarization \cite{phang2022investigating}, topic modeling \cite{yasunaga2019topiceq}, machine translation \cite{zhang2015deep}, speech recognition \cite{kamath2019deep}, lemmatization, part of speech tagging \cite{schmid1994part}, grammatical error correction \cite{dahlmeier2012better}, scholarly citation network analysis \cite{jha2017nlp}, named entity linking \cite{al2020named}, text to text and text to image generation etc \cite{manning1999foundations,bird2009natural,hirschberg2015advances,wolf2019huggingface}. Several web-tools such as Web of science, Scopus, Google scholar, Microsoft academic, Crossref, and PubMed etc. are using NLP to extract and analyze information from scholarly articles \cite{mongeon2016journal,shultz2007comparing,hendricks2020crossref,burnham2006scopus,wang2020microsoft} .

However, scientific literature, especially chemistry and materials science data, contains numerous technical terms (such as chemical names, methodologies, and instrumental techniques) that are difficult to process using conventional NLP tools. Luckily, there have been several advancements in applying NLP techniques to chemistry and materials science. One of the pioneer works for applying NLP for materials chemistry was carried out by Cole et al. \cite{swain2016chemdataextractor}, who demonstrated the application of NLP for magnetic and battery materials  \cite{court2018auto,huang2020database} using ChemDataExtractor. Other popular NLP for chemistry and materials tools include ChemicalTagger, ChemListem, ChemSpot, MaterialsParser, OSCAR4 details of which can be found elsewhere \cite{choudhary2022recent}. In addition to the magnetic and battery materials, similar works have been also performed for numerous other material classes such as metal organic frameworks \cite{nandy2022mofsimplify}, Mott insulator transition materials \cite{georgescu2021database}, glasses \cite{venugopal2021looking} and polymers \cite{shetty2021automated} etc. Other applications of NLP for materials data involve using Long short-term memory
 (LSTM) and transformer-based models to extract various categories of information, and in particular materials synthesis information from text sources \cite{weston2019named,vaucher2020automated,he2020similarity,kononova2019text,kim2017materials,tshitoyan2019unsupervised}. A detailed review of the application of NLP for materials can be found in refs. \cite{choudhary2022recent,olivetti2020data,kononova2021opportunities}.

Nevertheless, the application of NLP for materials applications is still an active area of development and there are a number of pain points that make NLP for materials data difficult. Some of these challenges are: 1) fully accessible text-data and pay-walls, 2) standard dataset and tools to benchmark NLP techniques for materials chemistry analysis, 3) example applications with both materials science and NLP domain knowledge, 4) resolving dependencies between words and phrases across multiple sentences and paragraphs and cross-domains, 5) examining the performance of text-generation applications (such as ChatGPT \cite{eloundou2023gpts} etc.) for chemistry applications.

In this work, we present ChemNLP library that 1) provides a curated and open-access arXiv and Pubchem datasets that can be directly used for NLP tasks, 2) share and illustrate software tools that can be used to visualize, analyse and perform various NLP tasks for materials chemistry specific text data, and 3) develop a user-interface to search chemistry data within available literature. Although in this work we primarily use arXiv and Pubchem dataset, the tools can be used for other infrastructures as well.

\section{Methods}

\subsection{Datasets}
In our work, we primarily use two open-access datasets: arXiv and PubChem. ArXiv is a collaboratively funded, community-supported resource and maintained and operated by Cornell University. In the present version, the arXiv has 1796911 pre-print articles collected over 34 years, hosting literature from scientific fields including Physics, Mathematics, and Computer Science and their sub-categories. Each pre-print in arXiv contains text, figures, authors, citations, categories, and other metadata which are of immense importance for NLP applications. These metadata are entered by users while uploading their manuscripts in arXiv. Similarly, PubChem dataset is an open-access collection of 43920 articles (in the version used here) including various topics such as deep-learning, covid-19, human-connectcome, brain-machine-interface, electroactive-polymer, pedot-electrodes and neuro-prosthetics. Originally both the datasets were obtained from the Kaggle competition and then converted into JARVIS-Tools data format \cite{choudhary2020joint} (\url{https://jarvis-tools.readthedocs.io/en/master/databases.html}). \red{Note that arXiv/pubhecm articles are often updated after peer-review process using corresponding version control process and that can have effects on the application of AI models. However, for the sake of reproducibility, we use a particular version of the dataset linked to FigShare in this work.  ChemNLP, methods can be applied on a later version of the dataset as well, if necessary.}


\subsection{WordCloud and Term frequency-inverse document frequency}

We visualize the frequency of one- and two-order n-grams using the WordCloud library \cite{mueller2018amueller}. Here, "n-" refers to consecutive words e.g. uni-grams (single words such as magnetic) and bi-grams (two consecutive words such as two-dimensional). 

Usually, machine learning algorithms require fixed-length input numerical vectors for supervised as well as unsupervised learning tasks. Some of the popular and simple method of feature extraction with text data are bag of words (BOW), term frequency-inverse document frequency (TF-IDF) and Word2Vec. We use TF-IDF in this work. TF-IDF is a numerical statistic that reflects importance of a word in a document. The TF-IDF value increases proportionally to the number of times a word appears in the document and is offset by the number of documents in the corpus that contain the word, which helps to adjust for the fact that some words appear more frequently in general. Term frequency (TF) indicates the significance of a particular term within the overall document. Term frequency, $TF(t,d)$, is the relative frequency of term $t$ within document $d$: 

\begin{equation} 
 TF(t,d)= \frac{f_{td}}{\sum_{t^{'} \in d} f_{t^{'}d}}
\end{equation} 

where $f_{td}$ is the raw count of a term in a document, i.e., the number of times that term $t$ occurs in document $d$. The denominator is simply the total number of terms in document $d$.

Hence, TF can be considered as the probability of finding a word in a document. The inverse document frequency (IDF) is a measure of how much information a word provides, i.e., if it’s common or rare across all documents. It is used to calculate the weight of rare words across all documents in the corpus. The words that occur rarely in the corpus have a high IDF score and vice versa. IDF is calculated as the logarithmically scaled inverse fraction of the documents that contain the word (obtained by dividing the total number of documents by the number of documents containing the term, and then taking the logarithm of that quotient):

\begin{equation} 
 IDF(t,D)= \log \frac{N}{|d \in D:t \in d|} 
\end{equation} 
where N is the total number texts/documents in the corpus $N=|D|$. Now, combining the above, the TFIDF is given as:
\begin{equation} 
 TFIDF(t,d,D)=  TF(t,d)*IDF(t,D)
\end{equation} 

\subsection{Clustering and text-classification}

 For clustering analysis, we use t-distributed stochastic neighbor embedding (t-SNE), which is a statistical method for visualizing high-dimensional data in a two- or three-dimensional map. The t-SNE plot was generated with the help of Scikit-learn library \cite{pedregosa2011scikit}. Note that clustering is an unsupervised ML task and the interpretations can be subjective. 
 
For the supervised machine learning classification task, we use various algorithms in the Scikit-learn library, graph neural networks as implemented using PyTorch \cite{paszke2019pytorch} and deep graph library (DGL) \cite{wang2019deep} and transformers using the huggingface library \cite{wolf2019huggingface}. 

There are a number of algorithms for supervised classification in Scikit-learn. We compare the results of four algorithms: support vector machine (SVM), random forest (RF) and logistic regression algorithms. The random forest algorithm is a type of supervised machine learning method based on ensemble learning. Ensemble learning is a based on joining different types of algorithms or same algorithm multiple times to form a more powerful prediction model. Support vector machine (SVM) finds a hyperplane in an N-dimensional space (where N is the number of features) that distinctly classifies the data points. Support vectors are data points that are closer to the hyperplane and influence the position and orientation of the hyperplane. Logistic Regression is a classification technique which uses a logistic function to model the dependent variable. 

We use TextGCN as proposed by Yao et al. \cite{yao2019graph} to convert text into graphs. TextGCN is initialized with one-hot representation for word and document. The edge between two word nodes is built by word co-occurrence information and the edge between a word node and document node is built using word frequency and word’s document frequency. We use graph neural network with attention layers (GAT) as the GNN model.

As a transformer model for text classification, we fine-tune the pre-trained \red{DistilBERT} model \cite{sanh2019distilbert}. DistilBERT is a small, fast, cheap and light transformer model based on the Bidirectional Encoder Representations from Transformers (BERT) architecture. Knowledge distillation is performed during the pre-training phase to reduce the size of a BERT model. DistilBERT leverages the inductive biases learned by larger models during pre-training, and uses a triple loss combining language modeling, distillation and cosine-distance losses.

We calculate the classification accuracy as:
\begin{equation}
    Accuracy=\frac{TP+TN}{TP + FN + FP + TN}
\end{equation}
where, TP, TN, FN, FP are True Positive, True Negative, False Negative, and False Positive instances respectively.

\subsection{Named entity recognition}
We use a pre-trained XLNet introduced by Yang et al. \cite{yang2019xlnet} and fine tune on the MatScholar NER dataset. \red{XLNet} is an extension of the Transformer-XL model pre-trained using an autoregressive method to learn bidirectional contexts by maximizing the expected likelihood over all permutations of the input sequence factorization order. XLNet is one of the few models that has no sequence length limit. XLNet also uses the same recurrence mechanism as Transformer-XL to build long-term dependencies. XLNet has recently been shown to outperform many oher models for NER tasks \cite{yan2021named}. The MatScholar dataset \cite{weston2019named} consists of the following tokens: material name (MAT), sample descriptor (DSC), symmetry/phase label (SPL), property (PRO), characterization method (CMT), application (APL) and synthesis method (SMT) for 800 manually annotated abstracts. NER typically uses BIO notation, which differentiates the beginning (B) and the inside (I) of entities. O is used for non-entity tokens.

The performance of the model is measured in terms of F1 score given by:

\begin{equation}
    F_1=\frac{2TP}{2TP + FP + TN}
\end{equation}
where, TP, TN, FN, FP are True Positive, True Negative, False Negative, and False Positive instances respectively. Similar to accuracy, the maximum value of F1 is 100 \%.

\subsection{Abstractive summarization and text-generation}

We use a pre-trained Text-to-Text Transfer Transformer (T5) \cite{raffel2020exploring} model developed by Google and then fine tune it to produce summaries of abstracts and analyze the performance by analyzing its closeness to the title of the given paper. Such summarizations are non-trivial tasks, but having title and abstract already available, its important to check whether the \red{large language models} (LLMs) can generate summary of abstracts that resemble corresponding titles. 

Note that unlike extractive summarization where works are extracted to make the summary, in abstractive summarizations, new words can be formed in order to make the gist of a text which is usually the case of titles of papers given their abstracts. T5 is a universal encoder-decoder model pre-trained on a multi-task mixture of unsupervised and supervised tasks and for which each task is converted into a text-to-text format. T5 works well on a variety of tasks out-of-the-box by prepending a different prefix to the input corresponding to each task, e.g., for translation, summarization, question answering, and classification. We use the t5-base model from the huggingface library, which has 220 million parameters. During the summarization process, it is important to append the text with a prefix "summarize:" in order for the model to understand the type of the task.

For the text-generation, we fine-tune the pre-trained Open Pre-trained Transformer (OPT) \cite{zhang2022opt} model developed by Meta AI. OPT was shown to match the performance and sizes of the generative pretrained transformers (GPT-3) class of models, while also applying the latest best practices \cite{schwartz2022towards} in data collection and efficient training. 

OPT was developed with an aim to enable reproducible and responsible research at scale for large language models (LLMs). OPT was predominantly pretrained with English text. The model was pretrained using a causal language modeling (CLM) objective. There are different types of OPT models with parameters ranging from 125 million to 175 billion. 

In this work, we choose to use the 125 million parameter model (facebook/opt-125m) for computational cost reasons. Note that similar OpenAI models such as GPT-3 and GPT-4 are not made public hence they cannot be easily finetuned and shared. Specifically, we fine tune the OPT model on cond-mat.supr-con category in the arXiv for the superconductor category (with 14697 articles) so that given titles, the model can generate abstracts. We follow the self-instruct \cite{wang2022self,alpaca} based instruction generation for converting the arXiv dataset before feeding into the model. Each entry in the dataset consists of three keys: 1) instruction, 2) input, 3) output. The instruction used here is "Describe the following:", the input is the title and output the abstract of a manuscript.


For the performance of these models, we use the Recall-Oriented Understudy for Gisting Evaluation (ROUGE) metric. ROUGE-N measures the overlap of n-grams between the generated text and the reference text. Here ROUGE-1 refers to overlap of unigram (each word), ROUGE-2 between  bigrams and so on. 

\subsection{Web-app}
We use ChemDataExtractor \cite{swain2016chemdataextractor} along with JARVIS-Tools and named entity recognition models to identify chemical names from abstracts in condensed matter category articles. ChemDataExtractor is a toolkit for the automated extraction of chemical entities and their associated properties, measurements, and relationships from scientific documents that can be used to populate structured chemical databases. Using the chemical information, we develop a web-app using JARVIS-Tools and Configurable Data Curation System (CDCS) which is primarily based on Django-python and java-script libraries. The web-app can be used to find articles given a chemical system. In future, we plan to share other models such as text classification, summarization, text generation, named entity recognition etc. using the same app as well which we believe would be of great use for materials research community. \red{As the datasets evolve, we plan to update the webapp as well.}

\subsection{Integration with DFT datasets}
The arXiv dataset consists of multiple materials classes such as superconductors, strongly correlated electron materials and so on. Finding overlaps and difference betwenn the exisiting literature and large scale databases can guide and expedite materials discovery. It is beyond the scope of this work to analyze NLP applications for each material class. However, we analyze the superconducting material information in the dataset and compare with recently developed JARVIS-SuperconDB dataset \cite{choudhary2022designing} to identify new and common chemical formulae. However, the dataset and tools presented here can be used for other material classes as well.

\section{Results and discussion}
\begin{figure}[hbt!]
    \centering
    \includegraphics[trim={0. 0cm 0 0cm},clip,width=0.98\textwidth]{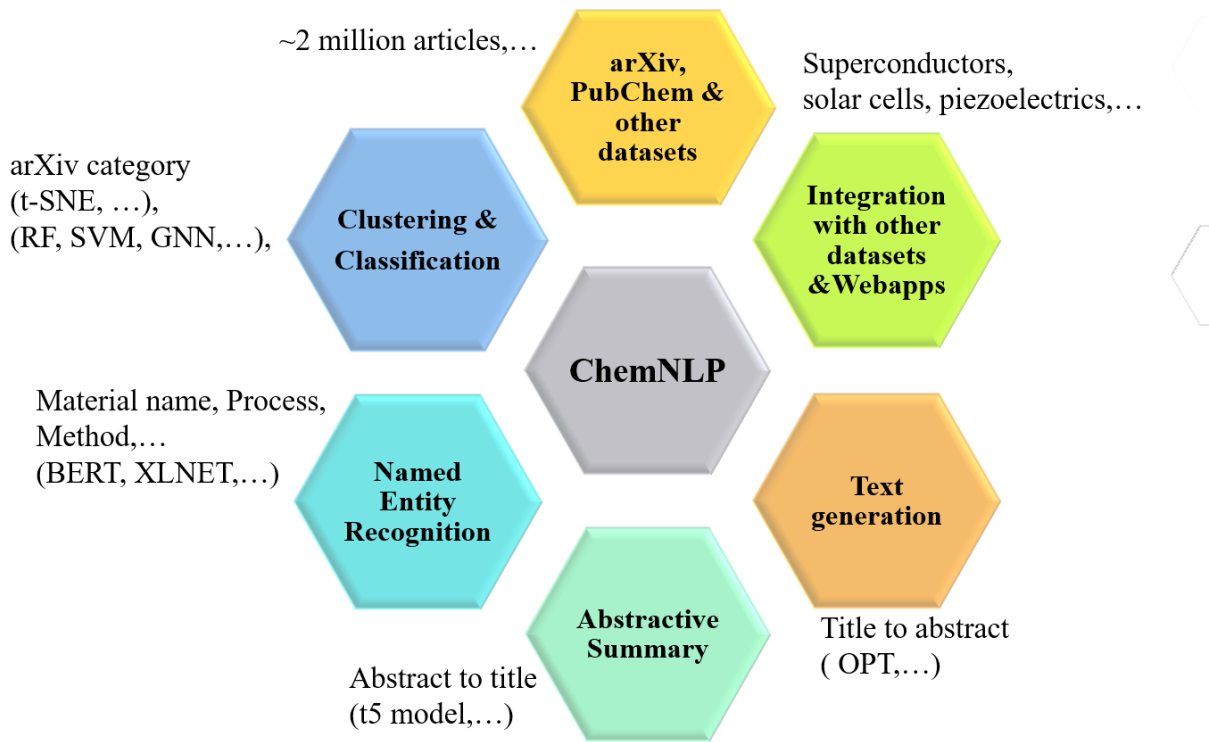}
    \caption{\label{fig:schematic}{A Schematic overview of the ChemNLP library. ChemNLP aims to provide a software toolkit with integrated dataset and comprehensive AI/ML tools for expanding natural language processing technique applications for tasks such as text classification, clustering, named entity recognition, abstractive summarization and text generation.}}
\end{figure}
A schematic overview of the ChemNLP library is shown in Fig.~\ref{fig:schematic}. ChemNLP serves as an open-access software package for a variety of natural language processing (NLP) tasks especially for chemistry and materials science applications. We discuss these applications and details in the following sections.

\subsection{Dataset analysis}

\begin{figure}[hbt!]
    \centering
    \includegraphics[trim={0. 0cm 0 0cm},width=1.0\textwidth]{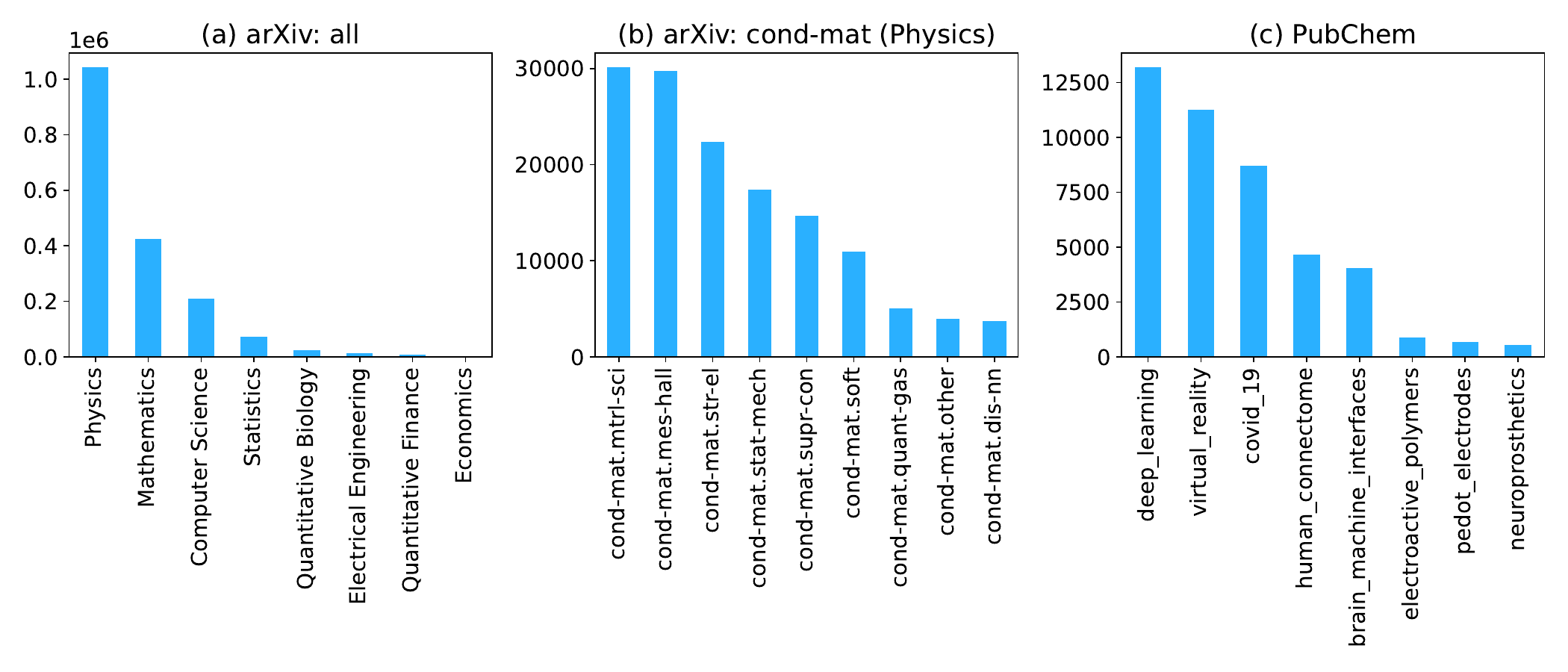}
    \caption{Several categories of scholarly articles in the arXiv dataset. a) overall categories, b) condensed matter sub-categories, c) PubChem dataset sub-vategories.\label{fig:categories}}
\end{figure}

The arXiv dataset provides a rich, multi-modal and large dataset for scientific literature. In Fig.~\ref{fig:categories}a, we show several categories of scholarly articles in the arXiv dataset. Such categorizations are possible because of 167 taxonomy categories data available in the dataset. The details of the taxonomy can be found at \url{https://arxiv.org/category_taxonomy}. In Fig.~\ref{fig:categories}a we notice that most of the articles belong to Physics, Mathematics and Computer science. The number of articles in Physics, Mathematics, Computer science, Statistics, Quantitative Biology, Electrical Engineering, Quantitative Finance and Economics are 1042227, 425745, 209068, 72058, 24720, 13064, 8920, and 1109 respectively.  Furthermore, we visualize the condensed matter physics categories in Fig.~\ref{fig:categories}b. The cond-mat.mtrl-sci, cond-mat.mes-hall, cond-mat.str-el, cond-mat.stat-mech, cond-mat.supr-con, cond-mat.soft, cond-mat.quant-gas, cond-mat.other, and cond-mat.dis-nn categories have 30107, 29751, 22375, 17359, 14697, 10939, 5041, 3930 and 3728 articles respectively.

Similarly, we show the PubChem dataset distribution in Fig.~\ref{fig:categories}c for 43920 
 articles. The sub-categories and their number counts (in parenthesis) available in PubChem are: deep-learning (13187), virtual-reality (11245), covid-19 (8694), humane-connectcome (4646), brain-machine-interface (4052), electroactive-polymer (896), pedot-electrodes (666), and neuroprosthetics (534).

\begin{figure}[hbt!]
    \centering
    \includegraphics[trim={0. 0cm 0 0cm},width=0.95\textwidth]{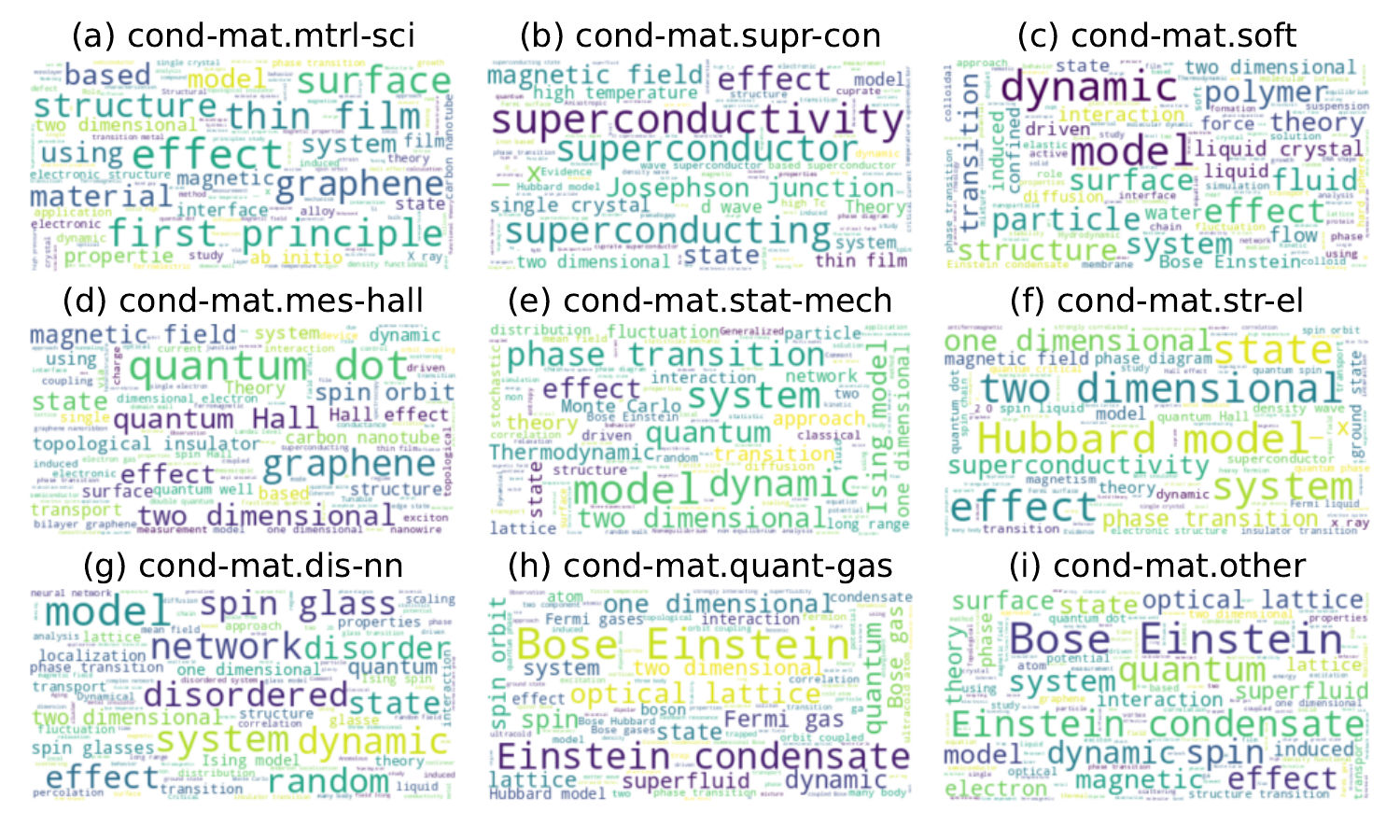}
    \caption{Word-cloud charts for different words in major condensed matter article titles. a) cond-mat.mtrl-sci, b) cond-mat.supr-con, c)cond-mat.soft, d) cond-mat.mes-hall, e) cond-mat.stat-mech, f) cond-mat.str-el, g) cond-mat.dis-nn, h) cond-mat.quant-gas, i) cond-mat.other\label{fig:ngram}}
\end{figure}

Next, in Fig.~\ref{fig:ngram} we show word-cloud charts for different words in the condensed matter articles' titles. A word cloud is a collection, or cluster, of words depicted in different sizes. The bigger and bolder the word appears, the more often it's mentioned within a given text and the more importance it holds. We find that first-principle, electronic structure, graphene, thin film, surface, carbon nanotube, two dimensional, magnetic etc. are some of the most common words in this subtopic. Similarly, Josephson junction, superconductivity, d wave, single crystal etc. are the most common words in the super-con category. Polymer, diffusion and fluid words are common ones in the soft category. Words like quantum-dot, quantum-well, spin-orbit, Hall-effect, topological insulator, quantum hall are common in mesh-hall category. The stat-mech category has phase-transition, fluctuation, dynamic, Thermodynamic etc. as some of the common words. The words like Hubbard-model, density-wave, spin-liquid etc. are common in str-el category. Words such as disorder, spin glass, network etc. are common in dis-nn category. Interestingly, some words such as two-dimensional occur in all the categories showing such class are one of the highly investigated materials. Similarly, the words such as magnetic and superconductivity occur in multiple categories showing such class of materials are investigated by experts in multiple domains. Similar analysis can be done for the PubChem dataset as well with the tools provided in the ChemNLP package.

\subsection{Clustering analysis}
\begin{figure}[hbt!]
    \centering
    \includegraphics[trim={0. 0cm 0 0cm},width=0.95\textwidth]{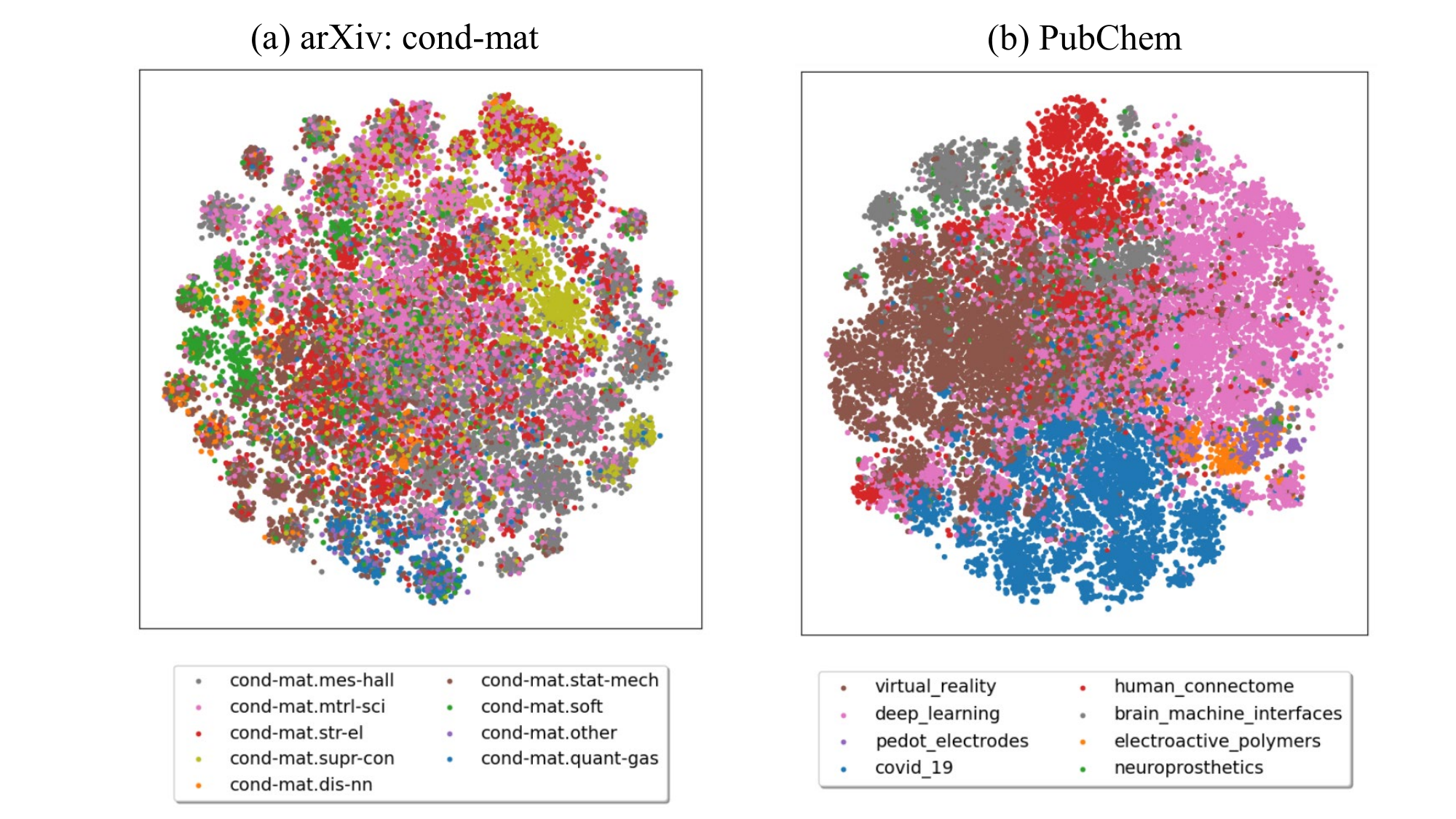}
    \caption{The t-SNE visualisations of the cond.mat articles in the arXiv and PubChem datasets. a) arXiv: cond-mat, b) PubChem.\label{fig:tsne}}
\end{figure}

In this section we apply t-distributed stochastic neighbor analysis (t-SNE) to visualize various categories in the arXiv and Pubchem datasets. Such analysis cannot be easily carried out with exploratory data analysis (EDA) carried out in the previous section. 

The t-SNE reveals local structure in high-dimensional data, placing points in the low-dimensional visualization close to each other with high probability if they have similar high-dimensional feature vectors. First, we stem the paper titles from the text corpus, and get "Term Frequency-Inverse Document Frequency" (TF-IDF) of a given word stem. The TF-IDF is a product of the {relative frequency with which a term appears in a single document} and the log of the total number of papers in the document pool divided by the number of documents in which a term appears. Then we perform truncated singular value decomposition (TruncatedSVD) for sparse data to reduce the dimensionality of the embedding space (128 size). Furthermore, we perform t-SNE to reduce embedding space to 2-dimension. 

The t-SNE plot thus obtained is shown in Fig.~\ref{fig:tsne}a with the marker colors indicate the article category of each article in arXiv:cond-mat category. We show the cluster of super-con, mes-hall, dis-nn and stat-mech while matrl-sci category seems to overlap with other classes as well. 

Next, we show the t-SNE plot for the PubChem dataset in Fig.~\ref{fig:tsne}b. Similar to the cond-mat articles, the t-SNE can cluster different categories of articles in the PubChem dataset as well. These plots suggests the article data is well-distributed i.e., not clustered and we can group them by converting the text in just two dimensions. 


\subsection{Text classification}

\begin{table}
\caption{Comparison of models for classifying articles.}\label{tab:classification}
\begin{tabular}{@{}llll@{}}
   \hline
   &&arXiv&\\
    \hline
Model & Title only & Abstract only & Title+Abstract\\
   \hline

SVM& 84.8& 90.8& 91.3\\
DistilBERT&86.7&89.9&90.0\\
RF&86.9 &88.4 &88.6 \\
LR& 79.2&85.0&86.0\\
GNN&76.0&81.0&82.0\\
   \hline
   &&PubChem&\\
    \hline
SVM&94.5&94.0&97.6\\
DistilBERT&94.5&97.0&97.5\\
RF&94.3&94.0&96.7\\
LR& 92.0&93.0&96.7\\
GNN&85.0&91.0&92.0\\
   \hline

\end{tabular}
\end{table}


While clustering analysis can be useful for qualitative interpretation, more quantitative analysis can be done with supervised classification training. We apply classification models for both the arXiv:cond-mat and PubChem datasets into their categories. ML algorithms cannot directly process text data so we convert the texts to numerical vectors using bag of words model and Term Frequency, Inverse Document Frequency (TF-IDF) as available in the Scikit-learn package. We choose 1) title, 2) abstracts and 3) titles along with abstracts text to classify the article as discussed above. After converting the text in the dataset to numerical representation, we apply a few well-known ML algorithms such as random-forest (RF), linear support vector machine \red{(SVM)},logistic regression (LR) and graph neural networks (GNN). We use a 80-20 \% split and show the performance in Table.~\ref{tab:classification}.

For all the models, we find that highest accuracy is achieved for title along with abstracts. Highest accuracy (91.3 \% for arXiv:cond-mat, and 97.6 \% for PubChem) are achieved with the SVM models mainly. As a baseline, the random guessing/baseline model has an accuracy of $1/9 = 11.11 \%$ (for nine classes) for arXiv:cond-mat and $1/8 = 12.5 \%$, hence the ML models are more than 6 times accurate than a random guessing models. The SVM model accuracy is followed by \red{DistilBERT} model, which has a slightly lower performance than SVM. Next, the models RF, LR and GNN models follow the \red{DistilBERT} model.




\begin{figure}[hbt!]
    \centering
    \includegraphics[trim={0. 0cm 0 0cm},width=0.95\textwidth]{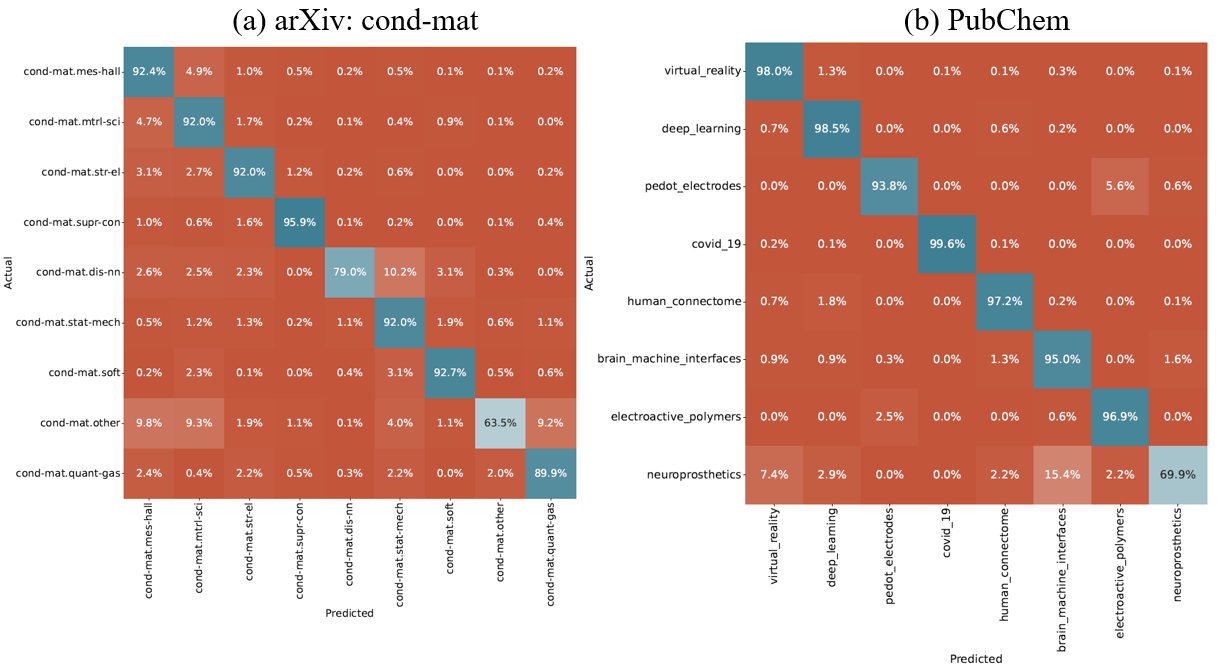}
    \caption{Confusion matrix for classifying cond-mat articles with linear support vector machine model. a) arXiv: cond-mat, b) PubChem.\label{fig:conf}}
\end{figure}

We show the classification confusion matrix for the combined title and abstract model for the two datasets in Fig.~\ref{fig:conf}. The confusion matrix allows us to interpret and analyze the detailed accuracy of the model for each class rather than just a global accuracy value. Ideally, a perfect classifier would result in a confusion matrix with diagonal entries only with a value of 100 \% . We find that a vast majority of the predictions end up on the diagonal (predicted label = actual label). This is especially true for the supr-con, mes-hall and matrl-sci subfields with the supr-con class showcasing the highest value at 95.9 \%, which is interesting. The lowest accuracy was achieved for the cond-mat.other subfield, which significantly overlaps with other categories such as mes-hall and quant-gas. The dis-nn sub-category also overlaps with stat-mech su-category. Beyond these two categories (other and dis-nn), we obtain close to or over 90 \% for all the sub-categories in the confusion matrix, demonstrating the promise of accurate classification through machine learning. Similarly, for the PubChem dataset, we achieve more that 93 \% accuracy except the neuroprosthetics category. This can also be attributed to their lower low counts as shown in Fig.~\ref{fig:categories}b and Fig.~\ref{fig:categories}c.

\subsection{Named Entity Recognition}

While overall classification of texts can be helpful for many applications, often word by word classification of texts allow rich mining of the text data. Named Entity Recognition (NER) or token classification is used as a text-mining approach for extracting meaningful information (called entities). An entity can be a word or a group of words such name, location, organization etc. For chemistry applications, they can be used for extracting information such as material name (MAT), sample descriptor (DSC), symmetry/phase label (SPL), property (PRO), characterization method (CMT), application (APL) and synthesis method (SMT). In this section we use the MatScholar dataset \cite{weston2019named} and train a transformer model with XLNet \cite{yan2021named} for extracting entities with high accuracy. After training the model, we apply the model to arXiv title, abstract as well as full texts to develop a database of entities relevant for materials design. The MatScholar dataset consists of 800 hand-annotated materials science abstracts to be used as training data. The data has an 80-10-10 split, giving 640 abstracts in the training set, 80 \% in the training set, and 10 \% in dev set and 10 \% in the test set.

\begin{figure}[hbt!]
    \centering
    \includegraphics[trim={0. 0cm 0 0cm},width=0.95\textwidth]{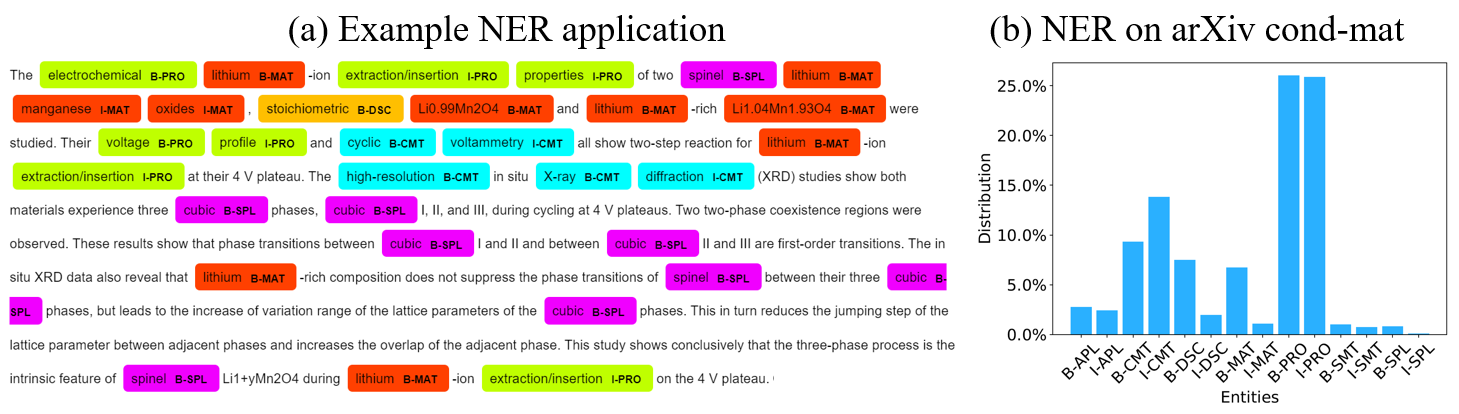}
    \caption{Named entity recognition/token classification for arXiv:cond-mat dataset. a) Example NER appplication. The tokens contain material name (MAT), sample descriptor (DSC), symmetry/phase label (SPL), property (PRO), characterization method (CMT), application (APL) and synthesis method (SMT). a) in this example, the red color represents material name (MAT), green the property (PRO), cyan the characterization method (CMT), orange the sample descriptor (DSC), pink the symmetry/phase labels (SPL). b) distribution of various entities in cond-mat arXiv abstracts. \label{fig:ner}}
\end{figure}

XLNet is a BERT-like transformer model and is known to outperform BERT substantially. Unlike BERT, XLNet is an autoregressive (AR) language model, which uses the context word to predict the next word. Here the context word is constrained to two directions either forward or backward. Moreover, we use a batch size of 64, 50 epochs, 5e-5 learning rate, max sequence length of 128 with AdamW optimizer. The F1 accuracies on the development and test sets are 87 \% which are reasonable and comparable to the previous work \cite{weston2019named}. An example of NER application on an abstract from ref. \cite{sun2002situ} is shown in Fig.~\ref{fig:ner}a. Here several types of text tokens are highlighted by different colors. The red color represents material name (MAT), green the property (PRO), cyan the characterization method (CMT), orange the sample descriptor (DSC), pink the symmetry/phase labels (SPL). Moreover, we show the entity distribution in the arXiv cond-mat articles in Fig.~\ref{fig:ner}b. We find that abstracts are highly dominated by property tags and least with materials specific names.

\subsection{Text-to-Text models: abstractive summarization and text-generation}

\begin{figure}[hbt!]
    \centering
    \includegraphics[trim={0. 0cm 0 0cm},width=0.95\textwidth]{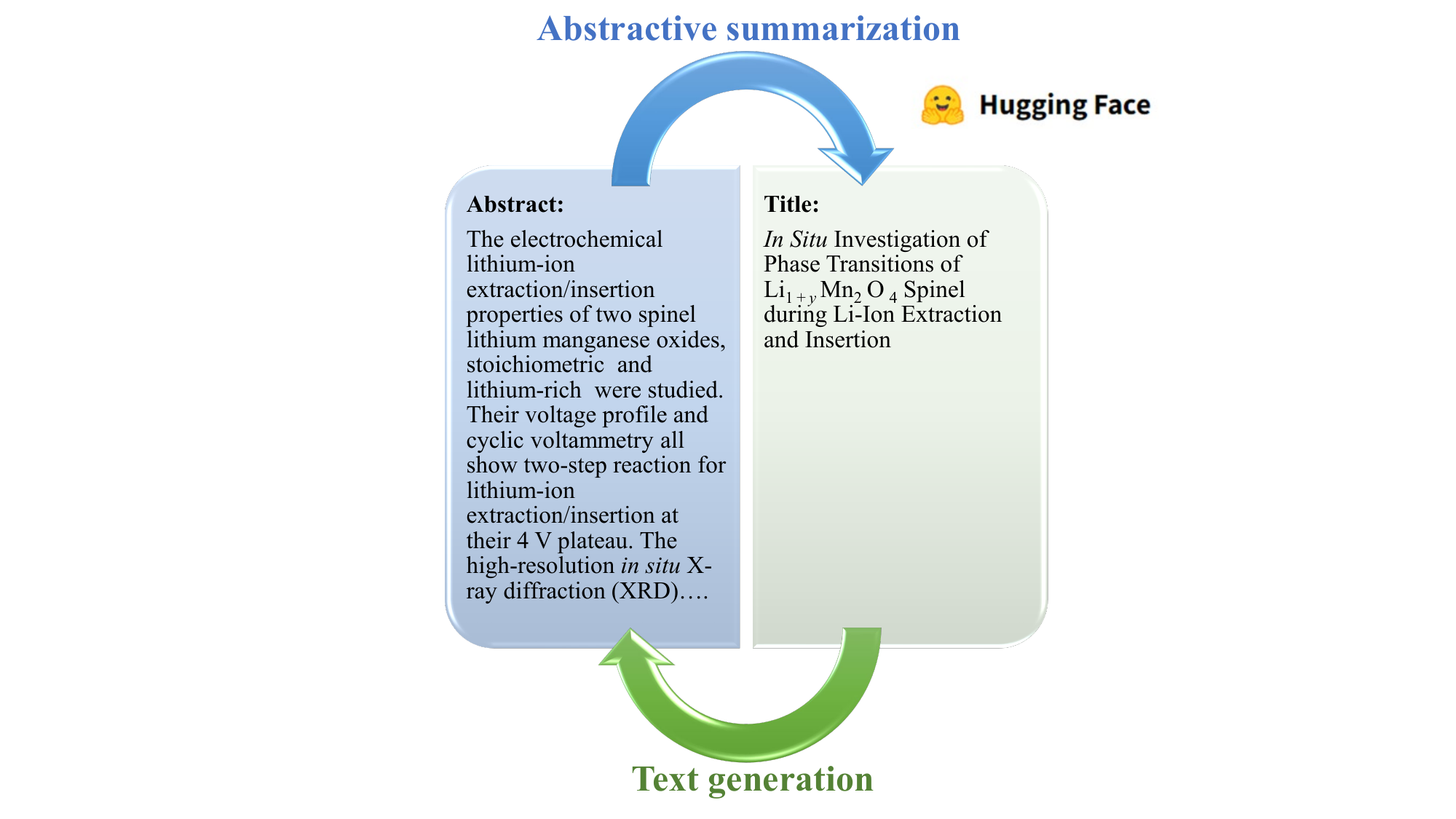}
    \caption{Text generation models for abstractive summarization (e.g., with Text-To-Text Transfer Transformer (T5) model) and text generation (e.g., with Open Pre-trained Transformer (OPT) model). For this exercise, we use the title and abstract data available in the arXiv-cond-mat dataset. Here, the titles are considered as summaries and abstracts are considered as generated texts.\label{fig:gen}}
\end{figure}

Text to text generation is one of the most rapidly developing field in natural language processing especially after the revolution in large language model (LLM) field. There are many tasks for text-to-text generation but here we choose to focus on two applications: 1) abstract summarizations e.g., making titles from the manuscript abstracts, 2) text generation e.g., generating abstracts from the titles of the papers as shown by an example in Fig.~\ref{fig:gen}. Obviously, there is not a unique approach for both the tasks. Nevertheless, such abstracts and title are generated by highly qualified scientists and engineers and can act as a decent benchmark for comparisons. Moreover, developing metrics for text-generation is  a difficult task. Here, for sake of simplicity we choose the ROUGUE metrics for evaluating the model performance.

Summarization tasks are of two types: extractive summary (the model extracts the most important sentences from the article and puts them together) and abstractive summary (the model creates new sentences to encapsulate maximum gist of the article). We obtain a pretrained unified t5 (Text-to-Text Transfer Transformer) model from Raffel et al. \cite{raffel2020exploring} and fine tune it for the arXiv dataset. Unlike BERT-style models that can only output either a class label or a span of the input, T5 propose reframing all NLP tasks into a unified text-to-text-format where the input and output are always text strings. T5 is an encoder-decoder model pre-trained on a multi-task mixture of unsupervised and supervised tasks. T5 has been used for machine translation, document summarization, question answering, and classification tasks (e.g., sentiment analysis) and hence provide a unified framework for NLP.  

We use 80:20 train-test split of arXiv-cond-mat dataset and evaluate the model performance of the test set using ROUGE (Recall-Oriented Understudy for Gisting Evaluation) score. So, the training dataset has 110342 entries while the test dataset has 27585 entries. We use a prefix of "summarize:" in front of the abstracts to instruct the model for the summarization task. After training, we found a ROGUE-1 score of 46.5 \% which is much better than an untrained model score of 30.8 \% i.e., a model without being exposed to the cond-mat data in the training set. Clearly, fine-tuning the model can help improve the performance. The ROGUE score obtained here is similar to scores achieved in other tasks for abstractive summarization (usually under 50 \%)\cite{xiao2021primer}.

Similarly, we fine-tune a pre-trained OPT model to generate abstract given the title of an article. The prefix used during the generation process was: "Describe the following:", similar to the "summarize:" prefix for the summarization task. We can also use the suffix "can be described as" for the title.s We use a similar 80 \%-20 \% train-test split for the data. After fine-tuning we find a ROGUE score of 37 \% which is reasonable for pre-suggestion applications. The trained model can now be used to generate text given a scientific idea such as a title. Ofcourse, there are many other text-generative models that are being developed and it will be very interesting to see how the quality of generated texts evolve as the model quality improves. Hence, ChemNLP has a flexible format to fine tune the models that may come in future.

\begin{figure}[hbt!]
    \centering
    \includegraphics[trim={0. 0cm 0 0cm},width=0.95\textwidth]{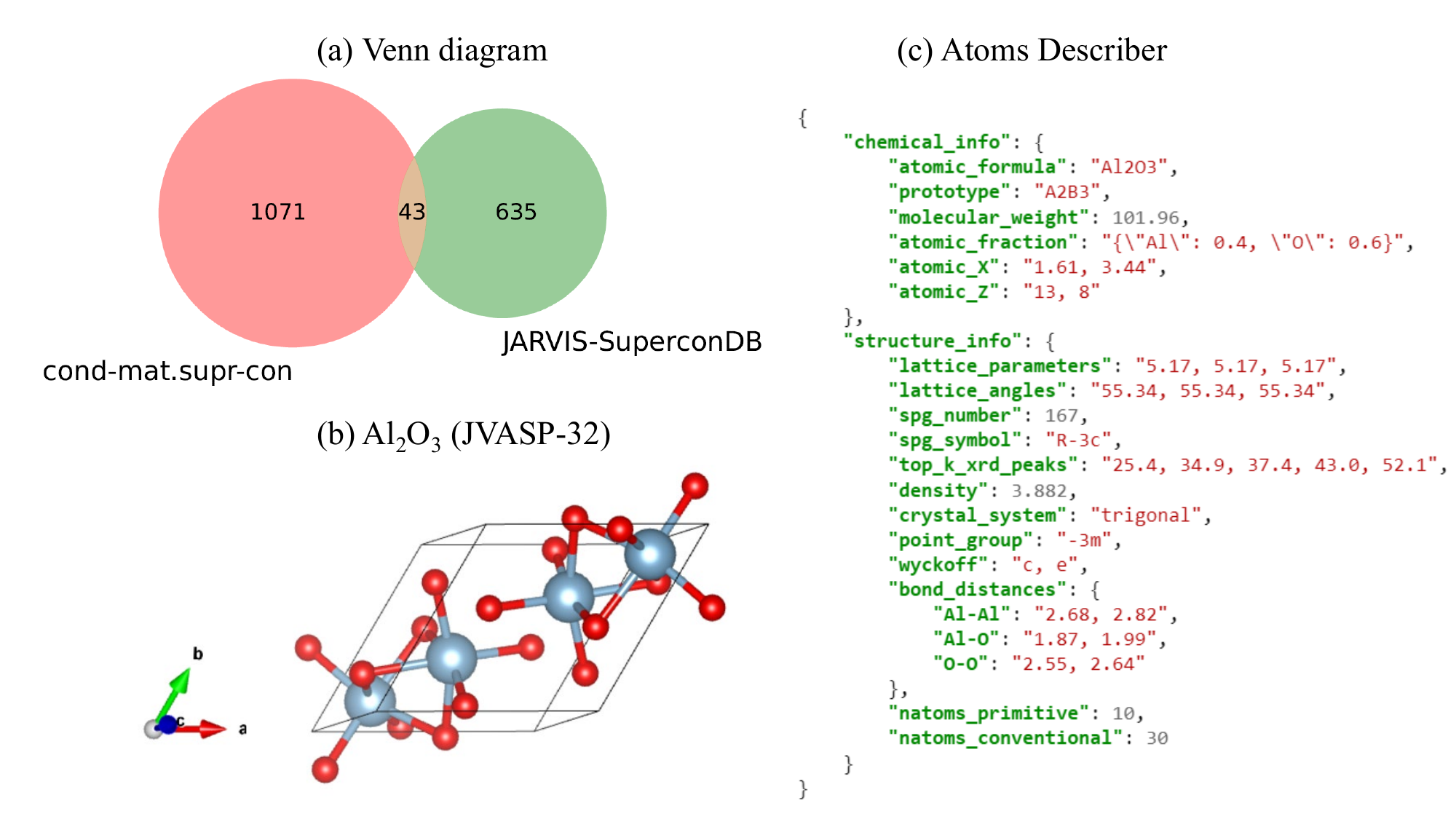}
    \caption{Integration with DFT databases. a) Venn diagram for chemical formula available in arXiv cond-mat.supr-con and JARVIS-SuperconDB. Out of \~1058 materials currently available in the JARVIS-SuperconDB, only 43 were found in the arXiv super-con condensed matter physics abstracts suggesting disovery of new superonductors and also lack of unconventional superconductors in the DFT databases. b) An example atomic structure for aluminum oxide, c) ChemNLP atomic describer example output. \label{fig:venn}}
\end{figure}

\subsection{Integration with DFT database}

The ChemNLP allows an easy integration of arXiv \red{dataset} information with DFT datasets such as JARVIS-DFT \cite{choudhary2020joint}. Such datasets contains various computed properties such as atomic structures, formation energies, bandgaps, elastic, thermoelectric, piezoelectric, dielectric, magnetic, solar cell, vibrational and superconducting properties. Here we show an example application for the superconducting properties but similar studies can be carried out for other properties as well. We demonstrate a simple application of ChemNLP for discovery and design of superconductors. Superconductors are class of materials with vanishing electrical resistance under a characteristic temperature called the superconducting transition temperature. Superconductors can be both phonon and non-phonon mediated. Recently, we developed a phonon-mediated superconducting transition temperature database using density functional theory with more than 1000 materials (JARVIS-SuperconDB) \cite{choudhary2022designing} in the JARVIS-DFT database. The JARVIS-SuperconDB was based on the Debye temperature, electronic density of states at the Fermi-level, and subsequent McMillan-Allen-Dynes based formulation. We obtained the chemical formula from the articles in  cond-mat.supr-con abstracts and that from the JARVIS-SuperconDB and plot a Venn diagram in Fig.~\ref{fig:venn}a. Note that we compare based on chemical formula only ignoring the crystal structure information. Interestingly, we find only 43 materials common in these two sets including well-known materials such as MgB$_2$, Nb, NbN, HfN, Nb, Al, TiN, and VRu etc. There are 635 chemical formula with DFT $T_C \geq 1K$ which we didn't find in the arXiv dataset and 1071 formula were present in the arXiv dataset only. There are many unconventional/non-phonon mediated superconductors such as Yttrium barium copper oxide, and high pressure as well as doped superconductors such as \red{NaFe$_{1-x}$Co$_x$As} which are not present in current JARVIS-SuperconDB. Additionally, several novel superconductors predicted such as  KB$_6$, Ru$_3$NbC, V$_3$Pt, ScN, LaN$_2$ which are not available in literature to the best of our knowledge. Therefore, NLP combined with traditional materials design motivates us in our further screening of superconductors. In JARVIS-DFT, there are many other properties such as thermoelectric, magnetic, dielectric, piezoelectric, topological, elastic, thermodynamic, vibrational, nuclear and low-dimensional properties on which similar strategies could be applied but its currently beyond the scope of the present paper.

\red{ChemNLP can also be used with DFT databases to generate formatted description for atomic structure information. Such results can be used for training for example large language models. Some of the chemical and structural data available are chemical formula, molecular weight, lattice parameters, X-ray diffraction peaks, bond-distances and number of atoms. One of the common inputs for LLMs are javascript object notation(json)-formatted text. Here in Fig.~\ref{fig:venn}b, we show an example atomic structure of aluminum oxide (JARVIS-DFT identifier: JVASP-32), and its corresponding json formatted description in Fig.~\ref{fig:venn}c. ChemNLP is integrated with JARVIS-Tools package, so it can be used for including/excluding additional/redundant data as inputs to LLMs for a chemical formula or an atomic structure. In future, we plan to use such formatted json outputs for training material property prediction models such as for formation energies and bandgaps etc.}


\subsection{Webapp development}
\begin{figure}[hbt!]
    \centering
    \includegraphics[trim={0. 0cm 0 0cm},width=0.95\textwidth]{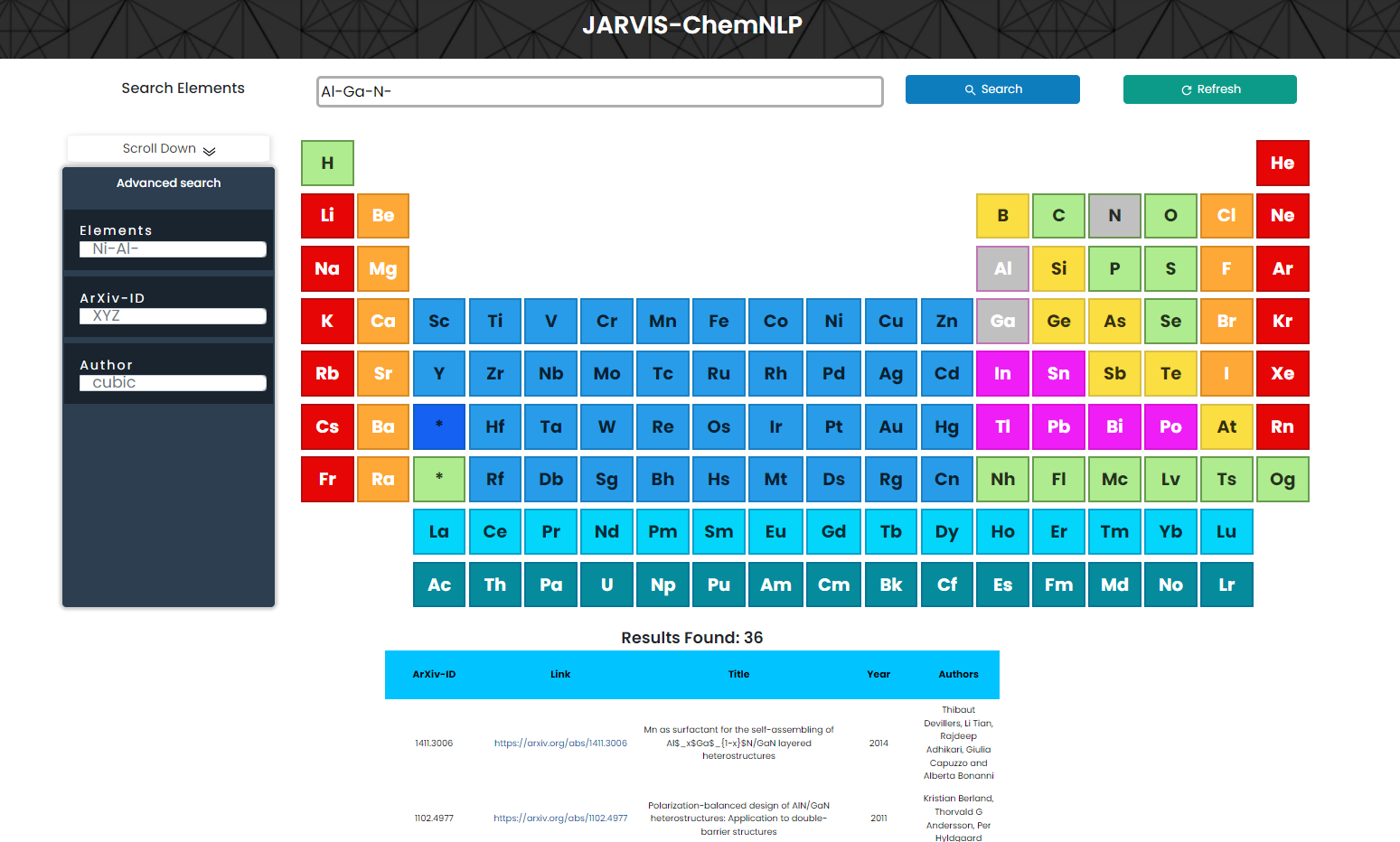}
    \caption{A snapshot of the web-app that can be used to find articles containing periodic table elements. This is based on arXiv condensed matter article abstracts. The web-app is available at : https://jarvis.nist.gov/jarvischemnlp.\label{fig:PTable}}
\end{figure}

Next, in Fig.~\ref{fig:PTable} we show a snapshot of the web-app that can be used to find articles containing periodic table elements. For instance, as we click on the elements Al, Ga and N and click Search button it returns 36 entries as results which can be used to find details of respective articles. Similarly, if element combinations such as Mo and S are selected, 1374 results are returned showing that articles with Mo-S compounds are much larger than Al-Ga-N. For this work, we collected all the abstracts in condensed matter Physics articles and attempted to find chemical formula with a combination of ChemDataExtractor and JARVIS-Tools packages. We found 37944 articles ($\sim$30 \% of condensed matter Physics entries) with a chemical formula in abstracts using the above approach. The number of unique combinations of elements (such as Al-Ga-N, Mo-Te, and Cu-Ga-In-S-Se etc.) that can be searched using the app are 6295.


In summary, we have developed a ChemNLP package (available at \url {https://github.com/usnistgov/chemnlp}) that can be used to analyze important materials chemistry information using the publicly-available datasets. To bridge the gap between the materials science/chemistry and natural language community, we have demonstrated several use-cases that could be useful for scientific community. These models include t-SNE, random forest, support vector machine, graph neural network, and transformers. We show that fine-tuning pre-trained models can help the accuracy of the models for chemistry related tasks. The web-app can be used as an easy interface for querying chemistry related data. We have also integrated big datasets of chemistry and materials science so that they can act as complementary to each other. As new models and datasets evolve, they can be easily integrated in the ChemNLP package. The models and data will be also integrated in a larger benchmarking project called JARVIS-Leaderboard (\url{https://pages.nist.gov/jarvis_leaderboard/}) for enhancing reproducibility and transparency. Moreover, we plan to extend ChemNLP to generate multi-modality projects by integrating with other projects in JARVIS such as atomistic vision (AtomVision) \cite{atomvision} and atomistic line graph neural network (ALIGNN) libraries \cite{choudhary2021atomistic} in the future.

\section{Conflict of interest}
The authors declare that they have no conflict of interest.
\begin{acknowledgement}
K.C. thanks Jacob Collard and Talapady Bhat at NIST for helpful discussion. Contributions from K.C. were supported by the financial assistance award 70NANB19H117 from the U.S. Department of Commerce, National Institute of Standards and Technology as well as Microsoft startup funds.

\end{acknowledgement}

\section{Data and software availability}
The code and the data sets used in this work are available at websites: \url{https://jarvis.nist.gov/jarvischemnlp/}, \url{https://github.com/usnistgov/chemnlp}, 
\url{https://jarvis-tools.readthedocs.io/en/master/databases.html}, and \url{https://doi.org/10.6084/m9.figshare.22351717}.



\bibliography{achemso-demo}

\providecommand{\latin}[1]{#1}
\makeatletter
\providecommand{\doi}
  {\begingroup\let\do\@makeother\dospecials
  \catcode`\{=1 \catcode`\}=2 \doi@aux}
\providecommand{\doi@aux}[1]{\endgroup\texttt{#1}}
\makeatother
\providecommand*\mcitethebibliography{\thebibliography}
\csname @ifundefined\endcsname{endmcitethebibliography}
  {\let\endmcitethebibliography\endthebibliography}{}
\begin{mcitethebibliography}{58}
\providecommand*\natexlab[1]{#1}
\providecommand*\mciteSetBstSublistMode[1]{}
\providecommand*\mciteSetBstMaxWidthForm[2]{}
\providecommand*\mciteBstWouldAddEndPuncttrue
  {\def\EndOfBibitem{\unskip.}}
\providecommand*\mciteBstWouldAddEndPunctfalse
  {\let\EndOfBibitem\relax}
\providecommand*\mciteSetBstMidEndSepPunct[3]{}
\providecommand*\mciteSetBstSublistLabelBeginEnd[3]{}
\providecommand*\EndOfBibitem{}
\mciteSetBstSublistMode{f}
\mciteSetBstMaxWidthForm{subitem}{(\alph{mcitesubitemcount})}
\mciteSetBstSublistLabelBeginEnd
  {\mcitemaxwidthsubitemform\space}
  {\relax}
  {\relax}

\bibitem[Jinha(2010)]{jinha2010article}
Jinha,~A.~E. Article 50 million: an estimate of the number of scholarly
  articles in existence. \emph{Learned publishing} \textbf{2010}, \emph{23},
  258--263\relax
\mciteBstWouldAddEndPuncttrue
\mciteSetBstMidEndSepPunct{\mcitedefaultmidpunct}
{\mcitedefaultendpunct}{\mcitedefaultseppunct}\relax
\EndOfBibitem
\bibitem[Khabsa and Giles(2014)Khabsa, and Giles]{khabsa2014number}
Khabsa,~M.; Giles,~C.~L. The number of scholarly documents on the public web.
  \emph{PloS one} \textbf{2014}, \emph{9}, e93949\relax
\mciteBstWouldAddEndPuncttrue
\mciteSetBstMidEndSepPunct{\mcitedefaultmidpunct}
{\mcitedefaultendpunct}{\mcitedefaultseppunct}\relax
\EndOfBibitem
\bibitem[Chowdhary(2020)]{chowdhary2020natural}
Chowdhary,~K. Natural language processing. \emph{Fundamentals of artificial
  intelligence} \textbf{2020}, 603--649\relax
\mciteBstWouldAddEndPuncttrue
\mciteSetBstMidEndSepPunct{\mcitedefaultmidpunct}
{\mcitedefaultendpunct}{\mcitedefaultseppunct}\relax
\EndOfBibitem
\bibitem[Mikolov \latin{et~al.}(2013)Mikolov, Sutskever, Chen, Corrado, and
  Dean]{mikolov2013distributed}
Mikolov,~T.; Sutskever,~I.; Chen,~K.; Corrado,~G.~S.; Dean,~J. Distributed
  representations of words and phrases and their compositionality.
  \emph{Advances in neural information processing systems} \textbf{2013},
  \emph{26}\relax
\mciteBstWouldAddEndPuncttrue
\mciteSetBstMidEndSepPunct{\mcitedefaultmidpunct}
{\mcitedefaultendpunct}{\mcitedefaultseppunct}\relax
\EndOfBibitem
\bibitem[Phang \latin{et~al.}(2022)Phang, Zhao, and
  Liu]{phang2022investigating}
Phang,~J.; Zhao,~Y.; Liu,~P.~J. Investigating Efficiently Extending
  Transformers for Long Input Summarization. \emph{arXiv preprint
  arXiv:2208.04347} \textbf{2022}, \relax
\mciteBstWouldAddEndPunctfalse
\mciteSetBstMidEndSepPunct{\mcitedefaultmidpunct}
{}{\mcitedefaultseppunct}\relax
\EndOfBibitem
\bibitem[Yasunaga and Lafferty(2019)Yasunaga, and
  Lafferty]{yasunaga2019topiceq}
Yasunaga,~M.; Lafferty,~J.~D. Topiceq: A joint topic and mathematical equation
  model for scientific texts. Proceedings of the AAAI conference on artificial
  intelligence. 2019; pp 7394--7401\relax
\mciteBstWouldAddEndPuncttrue
\mciteSetBstMidEndSepPunct{\mcitedefaultmidpunct}
{\mcitedefaultendpunct}{\mcitedefaultseppunct}\relax
\EndOfBibitem
\bibitem[Zhang and Zong(2015)Zhang, and Zong]{zhang2015deep}
Zhang,~J.; Zong,~C. Deep Neural Networks in Machine Translation: An Overview.
  \emph{IEEE Intell. Syst.} \textbf{2015}, \emph{30}, 16--25\relax
\mciteBstWouldAddEndPuncttrue
\mciteSetBstMidEndSepPunct{\mcitedefaultmidpunct}
{\mcitedefaultendpunct}{\mcitedefaultseppunct}\relax
\EndOfBibitem
\bibitem[Kamath \latin{et~al.}(2019)Kamath, Liu, and Whitaker]{kamath2019deep}
Kamath,~U.; Liu,~J.; Whitaker,~J. \emph{Deep learning for NLP and speech
  recognition}; Springer, 2019; Vol.~84\relax
\mciteBstWouldAddEndPuncttrue
\mciteSetBstMidEndSepPunct{\mcitedefaultmidpunct}
{\mcitedefaultendpunct}{\mcitedefaultseppunct}\relax
\EndOfBibitem
\bibitem[Schmid(1994)]{schmid1994part}
Schmid,~H. Part-of-speech tagging with neural networks. \emph{arXiv preprint
  cmp-lg/9410018} \textbf{1994}, \relax
\mciteBstWouldAddEndPunctfalse
\mciteSetBstMidEndSepPunct{\mcitedefaultmidpunct}
{}{\mcitedefaultseppunct}\relax
\EndOfBibitem
\bibitem[Dahlmeier and Ng(2012)Dahlmeier, and Ng]{dahlmeier2012better}
Dahlmeier,~D.; Ng,~H.~T. Better evaluation for grammatical error correction.
  Proceedings of the 2012 Conference of the North American Chapter of the
  Association for Computational Linguistics: Human Language Technologies. 2012;
  pp 568--572\relax
\mciteBstWouldAddEndPuncttrue
\mciteSetBstMidEndSepPunct{\mcitedefaultmidpunct}
{\mcitedefaultendpunct}{\mcitedefaultseppunct}\relax
\EndOfBibitem
\bibitem[Jha \latin{et~al.}(2017)Jha, Jbara, Qazvinian, and Radev]{jha2017nlp}
Jha,~R.; Jbara,~A.-A.; Qazvinian,~V.; Radev,~D.~R. NLP-driven citation analysis
  for scientometrics. \emph{Natural Language Engineering} \textbf{2017},
  \emph{23}, 93--130\relax
\mciteBstWouldAddEndPuncttrue
\mciteSetBstMidEndSepPunct{\mcitedefaultmidpunct}
{\mcitedefaultendpunct}{\mcitedefaultseppunct}\relax
\EndOfBibitem
\bibitem[Al-Moslmi \latin{et~al.}(2020)Al-Moslmi, Oca{\~n}a, Opdahl, and
  Veres]{al2020named}
Al-Moslmi,~T.; Oca{\~n}a,~M.~G.; Opdahl,~A.~L.; Veres,~C. Named entity
  extraction for knowledge graphs: A literature overview. \emph{IEEE Access}
  \textbf{2020}, \emph{8}, 32862--32881\relax
\mciteBstWouldAddEndPuncttrue
\mciteSetBstMidEndSepPunct{\mcitedefaultmidpunct}
{\mcitedefaultendpunct}{\mcitedefaultseppunct}\relax
\EndOfBibitem
\bibitem[Manning and Schutze(1999)Manning, and Schutze]{manning1999foundations}
Manning,~C.; Schutze,~H. \emph{Foundations of statistical natural language
  processing}; MIT press, 1999\relax
\mciteBstWouldAddEndPuncttrue
\mciteSetBstMidEndSepPunct{\mcitedefaultmidpunct}
{\mcitedefaultendpunct}{\mcitedefaultseppunct}\relax
\EndOfBibitem
\bibitem[Bird \latin{et~al.}(2009)Bird, Klein, and Loper]{bird2009natural}
Bird,~S.; Klein,~E.; Loper,~E. \emph{Natural language processing with Python:
  analyzing text with the natural language toolkit}; " O'Reilly Media, Inc.",
  2009\relax
\mciteBstWouldAddEndPuncttrue
\mciteSetBstMidEndSepPunct{\mcitedefaultmidpunct}
{\mcitedefaultendpunct}{\mcitedefaultseppunct}\relax
\EndOfBibitem
\bibitem[Hirschberg and Manning(2015)Hirschberg, and
  Manning]{hirschberg2015advances}
Hirschberg,~J.; Manning,~C.~D. Advances in natural language processing.
  \emph{Science} \textbf{2015}, \emph{349}, 261--266\relax
\mciteBstWouldAddEndPuncttrue
\mciteSetBstMidEndSepPunct{\mcitedefaultmidpunct}
{\mcitedefaultendpunct}{\mcitedefaultseppunct}\relax
\EndOfBibitem
\bibitem[Wolf \latin{et~al.}(2019)Wolf, Debut, Sanh, Chaumond, Delangue, Moi,
  Cistac, Rault, Louf, Funtowicz, \latin{et~al.} others]{wolf2019huggingface}
Wolf,~T.; Debut,~L.; Sanh,~V.; Chaumond,~J.; Delangue,~C.; Moi,~A.; Cistac,~P.;
  Rault,~T.; Louf,~R.; Funtowicz,~M., \latin{et~al.}  Huggingface's
  transformers: State-of-the-art natural language processing. \emph{arXiv
  preprint arXiv:1910.03771} \textbf{2019}, \relax
\mciteBstWouldAddEndPunctfalse
\mciteSetBstMidEndSepPunct{\mcitedefaultmidpunct}
{}{\mcitedefaultseppunct}\relax
\EndOfBibitem
\bibitem[Mongeon and Paul-Hus(2016)Mongeon, and Paul-Hus]{mongeon2016journal}
Mongeon,~P.; Paul-Hus,~A. The journal coverage of Web of Science and Scopus: a
  comparative analysis. \emph{Scientometrics} \textbf{2016}, \emph{106},
  213--228\relax
\mciteBstWouldAddEndPuncttrue
\mciteSetBstMidEndSepPunct{\mcitedefaultmidpunct}
{\mcitedefaultendpunct}{\mcitedefaultseppunct}\relax
\EndOfBibitem
\bibitem[Shultz(2007)]{shultz2007comparing}
Shultz,~M. Comparing test searches in PubMed and Google Scholar. \emph{Journal
  of the Medical Library Association: JMLA} \textbf{2007}, \emph{95}, 442\relax
\mciteBstWouldAddEndPuncttrue
\mciteSetBstMidEndSepPunct{\mcitedefaultmidpunct}
{\mcitedefaultendpunct}{\mcitedefaultseppunct}\relax
\EndOfBibitem
\bibitem[Hendricks \latin{et~al.}(2020)Hendricks, Tkaczyk, Lin, and
  Feeney]{hendricks2020crossref}
Hendricks,~G.; Tkaczyk,~D.; Lin,~J.; Feeney,~P. Crossref: The sustainable
  source of community-owned scholarly metadata. \emph{Quantitative Science
  Studies} \textbf{2020}, \emph{1}, 414--427\relax
\mciteBstWouldAddEndPuncttrue
\mciteSetBstMidEndSepPunct{\mcitedefaultmidpunct}
{\mcitedefaultendpunct}{\mcitedefaultseppunct}\relax
\EndOfBibitem
\bibitem[Burnham(2006)]{burnham2006scopus}
Burnham,~J.~F. Scopus database: a review. \emph{Biomedical digital libraries}
  \textbf{2006}, \emph{3}, 1--8\relax
\mciteBstWouldAddEndPuncttrue
\mciteSetBstMidEndSepPunct{\mcitedefaultmidpunct}
{\mcitedefaultendpunct}{\mcitedefaultseppunct}\relax
\EndOfBibitem
\bibitem[Wang \latin{et~al.}(2020)Wang, Shen, Huang, Wu, Dong, and
  Kanakia]{wang2020microsoft}
Wang,~K.; Shen,~Z.; Huang,~C.; Wu,~C.-H.; Dong,~Y.; Kanakia,~A. Microsoft
  academic graph: When experts are not enough. \emph{Quantitative Science
  Studies} \textbf{2020}, \emph{1}, 396--413\relax
\mciteBstWouldAddEndPuncttrue
\mciteSetBstMidEndSepPunct{\mcitedefaultmidpunct}
{\mcitedefaultendpunct}{\mcitedefaultseppunct}\relax
\EndOfBibitem
\bibitem[Swain and Cole(2016)Swain, and Cole]{swain2016chemdataextractor}
Swain,~M.~C.; Cole,~J.~M. ChemDataExtractor: a toolkit for automated extraction
  of chemical information from the scientific literature. \emph{Journal of
  chemical information and modeling} \textbf{2016}, \emph{56}, 1894--1904\relax
\mciteBstWouldAddEndPuncttrue
\mciteSetBstMidEndSepPunct{\mcitedefaultmidpunct}
{\mcitedefaultendpunct}{\mcitedefaultseppunct}\relax
\EndOfBibitem
\bibitem[Court and Cole(2018)Court, and Cole]{court2018auto}
Court,~C.~J.; Cole,~J.~M. Auto-generated materials database of Curie and
  N{\'e}el temperatures via semi-supervised relationship extraction.
  \emph{Scientific data} \textbf{2018}, \emph{5}, 1--12\relax
\mciteBstWouldAddEndPuncttrue
\mciteSetBstMidEndSepPunct{\mcitedefaultmidpunct}
{\mcitedefaultendpunct}{\mcitedefaultseppunct}\relax
\EndOfBibitem
\bibitem[Huang and Cole(2020)Huang, and Cole]{huang2020database}
Huang,~S.; Cole,~J.~M. A database of battery materials auto-generated using
  ChemDataExtractor. \emph{Scientific Data} \textbf{2020}, \emph{7},
  1--13\relax
\mciteBstWouldAddEndPuncttrue
\mciteSetBstMidEndSepPunct{\mcitedefaultmidpunct}
{\mcitedefaultendpunct}{\mcitedefaultseppunct}\relax
\EndOfBibitem
\bibitem[Choudhary \latin{et~al.}(2022)Choudhary, DeCost, Chen, Jain, Tavazza,
  Cohn, Park, Choudhary, Agrawal, Billinge, \latin{et~al.}
  others]{choudhary2022recent}
Choudhary,~K.; DeCost,~B.; Chen,~C.; Jain,~A.; Tavazza,~F.; Cohn,~R.;
  Park,~C.~W.; Choudhary,~A.; Agrawal,~A.; Billinge,~S.~J., \latin{et~al.}
  Recent advances and applications of deep learning methods in materials
  science. \emph{npj Computational Materials} \textbf{2022}, \emph{8},
  1--26\relax
\mciteBstWouldAddEndPuncttrue
\mciteSetBstMidEndSepPunct{\mcitedefaultmidpunct}
{\mcitedefaultendpunct}{\mcitedefaultseppunct}\relax
\EndOfBibitem
\bibitem[Nandy \latin{et~al.}(2022)Nandy, Terrones, Arunachalam, Duan, Kastner,
  and Kulik]{nandy2022mofsimplify}
Nandy,~A.; Terrones,~G.; Arunachalam,~N.; Duan,~C.; Kastner,~D.~W.;
  Kulik,~H.~J. MOFSimplify, machine learning models with extracted stability
  data of three thousand metal--organic frameworks. \emph{Scientific Data}
  \textbf{2022}, \emph{9}, 1--11\relax
\mciteBstWouldAddEndPuncttrue
\mciteSetBstMidEndSepPunct{\mcitedefaultmidpunct}
{\mcitedefaultendpunct}{\mcitedefaultseppunct}\relax
\EndOfBibitem
\bibitem[Georgescu \latin{et~al.}(2021)Georgescu, Ren, Toland, Zhang, Miller,
  Apley, Olivetti, Wagner, and Rondinelli]{georgescu2021database}
Georgescu,~A.~B.; Ren,~P.; Toland,~A.~R.; Zhang,~S.; Miller,~K.~D.;
  Apley,~D.~W.; Olivetti,~E.~A.; Wagner,~N.; Rondinelli,~J.~M. Database,
  Features, and Machine Learning Model to Identify Thermally Driven
  Metal--Insulator Transition Compounds. \emph{Chemistry of Materials}
  \textbf{2021}, \emph{33}, 5591--5605\relax
\mciteBstWouldAddEndPuncttrue
\mciteSetBstMidEndSepPunct{\mcitedefaultmidpunct}
{\mcitedefaultendpunct}{\mcitedefaultseppunct}\relax
\EndOfBibitem
\bibitem[Venugopal \latin{et~al.}(2021)Venugopal, Sahoo, Zaki, Agarwal,
  Gosvami, and Krishnan]{venugopal2021looking}
Venugopal,~V.; Sahoo,~S.; Zaki,~M.; Agarwal,~M.; Gosvami,~N.~N.;
  Krishnan,~N.~A. Looking through glass: Knowledge discovery from materials
  science literature using natural language processing. \emph{Patterns}
  \textbf{2021}, \emph{2}, 100290\relax
\mciteBstWouldAddEndPuncttrue
\mciteSetBstMidEndSepPunct{\mcitedefaultmidpunct}
{\mcitedefaultendpunct}{\mcitedefaultseppunct}\relax
\EndOfBibitem
\bibitem[Shetty and Ramprasad(2021)Shetty, and Ramprasad]{shetty2021automated}
Shetty,~P.; Ramprasad,~R. Automated knowledge extraction from polymer
  literature using natural language processing. \emph{Iscience} \textbf{2021},
  \emph{24}, 101922\relax
\mciteBstWouldAddEndPuncttrue
\mciteSetBstMidEndSepPunct{\mcitedefaultmidpunct}
{\mcitedefaultendpunct}{\mcitedefaultseppunct}\relax
\EndOfBibitem
\bibitem[Weston \latin{et~al.}(2019)Weston, Tshitoyan, Dagdelen, Kononova,
  Trewartha, Persson, Ceder, and Jain]{weston2019named}
Weston,~L.; Tshitoyan,~V.; Dagdelen,~J.; Kononova,~O.; Trewartha,~A.;
  Persson,~K.~A.; Ceder,~G.; Jain,~A. Named entity recognition and
  normalization applied to large-scale information extraction from the
  materials science literature. \emph{Journal of chemical information and
  modeling} \textbf{2019}, \emph{59}, 3692--3702\relax
\mciteBstWouldAddEndPuncttrue
\mciteSetBstMidEndSepPunct{\mcitedefaultmidpunct}
{\mcitedefaultendpunct}{\mcitedefaultseppunct}\relax
\EndOfBibitem
\bibitem[Vaucher \latin{et~al.}(2020)Vaucher, Zipoli, Geluykens, Nair,
  Schwaller, and Laino]{vaucher2020automated}
Vaucher,~A.~C.; Zipoli,~F.; Geluykens,~J.; Nair,~V.~H.; Schwaller,~P.;
  Laino,~T. Automated extraction of chemical synthesis actions from
  experimental procedures. \emph{Nature communications} \textbf{2020},
  \emph{11}, 1--11\relax
\mciteBstWouldAddEndPuncttrue
\mciteSetBstMidEndSepPunct{\mcitedefaultmidpunct}
{\mcitedefaultendpunct}{\mcitedefaultseppunct}\relax
\EndOfBibitem
\bibitem[He \latin{et~al.}(2020)He, Sun, Huo, Kononova, Rong, Tshitoyan,
  Botari, and Ceder]{he2020similarity}
He,~T.; Sun,~W.; Huo,~H.; Kononova,~O.; Rong,~Z.; Tshitoyan,~V.; Botari,~T.;
  Ceder,~G. Similarity of precursors in solid-state synthesis as text-mined
  from scientific literature. \emph{Chemistry of Materials} \textbf{2020},
  \emph{32}, 7861--7873\relax
\mciteBstWouldAddEndPuncttrue
\mciteSetBstMidEndSepPunct{\mcitedefaultmidpunct}
{\mcitedefaultendpunct}{\mcitedefaultseppunct}\relax
\EndOfBibitem
\bibitem[Kononova \latin{et~al.}(2019)Kononova, Huo, He, Rong, Botari, Sun,
  Tshitoyan, and Ceder]{kononova2019text}
Kononova,~O.; Huo,~H.; He,~T.; Rong,~Z.; Botari,~T.; Sun,~W.; Tshitoyan,~V.;
  Ceder,~G. Text-mined dataset of inorganic materials synthesis recipes.
  \emph{Scientific data} \textbf{2019}, \emph{6}, 1--11\relax
\mciteBstWouldAddEndPuncttrue
\mciteSetBstMidEndSepPunct{\mcitedefaultmidpunct}
{\mcitedefaultendpunct}{\mcitedefaultseppunct}\relax
\EndOfBibitem
\bibitem[Kim \latin{et~al.}(2017)Kim, Huang, Saunders, McCallum, Ceder, and
  Olivetti]{kim2017materials}
Kim,~E.; Huang,~K.; Saunders,~A.; McCallum,~A.; Ceder,~G.; Olivetti,~E.
  Materials synthesis insights from scientific literature via text extraction
  and machine learning. \emph{Chemistry of Materials} \textbf{2017}, \emph{29},
  9436--9444\relax
\mciteBstWouldAddEndPuncttrue
\mciteSetBstMidEndSepPunct{\mcitedefaultmidpunct}
{\mcitedefaultendpunct}{\mcitedefaultseppunct}\relax
\EndOfBibitem
\bibitem[Tshitoyan \latin{et~al.}(2019)Tshitoyan, Dagdelen, Weston, Dunn, Rong,
  Kononova, Persson, Ceder, and Jain]{tshitoyan2019unsupervised}
Tshitoyan,~V.; Dagdelen,~J.; Weston,~L.; Dunn,~A.; Rong,~Z.; Kononova,~O.;
  Persson,~K.~A.; Ceder,~G.; Jain,~A. Unsupervised word embeddings capture
  latent knowledge from materials science literature. \emph{Nature}
  \textbf{2019}, \emph{571}, 95--98\relax
\mciteBstWouldAddEndPuncttrue
\mciteSetBstMidEndSepPunct{\mcitedefaultmidpunct}
{\mcitedefaultendpunct}{\mcitedefaultseppunct}\relax
\EndOfBibitem
\bibitem[Olivetti \latin{et~al.}(2020)Olivetti, Cole, Kim, Kononova, Ceder,
  Han, and Hiszpanski]{olivetti2020data}
Olivetti,~E.~A.; Cole,~J.~M.; Kim,~E.; Kononova,~O.; Ceder,~G.; Han,~T. Y.-J.;
  Hiszpanski,~A.~M. Data-driven materials research enabled by natural language
  processing and information extraction. \emph{Applied Physics Reviews}
  \textbf{2020}, \emph{7}, 041317\relax
\mciteBstWouldAddEndPuncttrue
\mciteSetBstMidEndSepPunct{\mcitedefaultmidpunct}
{\mcitedefaultendpunct}{\mcitedefaultseppunct}\relax
\EndOfBibitem
\bibitem[Kononova \latin{et~al.}(2021)Kononova, He, Huo, Trewartha, Olivetti,
  and Ceder]{kononova2021opportunities}
Kononova,~O.; He,~T.; Huo,~H.; Trewartha,~A.; Olivetti,~E.~A.; Ceder,~G.
  Opportunities and challenges of text mining in materials research.
  \emph{Iscience} \textbf{2021}, \emph{24}, 102155\relax
\mciteBstWouldAddEndPuncttrue
\mciteSetBstMidEndSepPunct{\mcitedefaultmidpunct}
{\mcitedefaultendpunct}{\mcitedefaultseppunct}\relax
\EndOfBibitem
\bibitem[Eloundou \latin{et~al.}(2023)Eloundou, Manning, Mishkin, and
  Rock]{eloundou2023gpts}
Eloundou,~T.; Manning,~S.; Mishkin,~P.; Rock,~D. GPTs are GPTs: An Early Look
  at the Labor Market Impact Potential of Large Language Models. \emph{arXiv
  preprint arXiv:2303.10130} \textbf{2023}, \relax
\mciteBstWouldAddEndPunctfalse
\mciteSetBstMidEndSepPunct{\mcitedefaultmidpunct}
{}{\mcitedefaultseppunct}\relax
\EndOfBibitem
\bibitem[Choudhary \latin{et~al.}(2020)Choudhary, Garrity, Reid, DeCost,
  Biacchi, Hight~Walker, Trautt, Hattrick-Simpers, Kusne, Centrone,
  \latin{et~al.} others]{choudhary2020joint}
Choudhary,~K.; Garrity,~K.~F.; Reid,~A.~C.; DeCost,~B.; Biacchi,~A.~J.;
  Hight~Walker,~A.~R.; Trautt,~Z.; Hattrick-Simpers,~J.; Kusne,~A.~G.;
  Centrone,~A., \latin{et~al.}  The joint automated repository for various
  integrated simulations (JARVIS) for data-driven materials design. \emph{npj
  Computational Materials} \textbf{2020}, \emph{6}, 1--13\relax
\mciteBstWouldAddEndPuncttrue
\mciteSetBstMidEndSepPunct{\mcitedefaultmidpunct}
{\mcitedefaultendpunct}{\mcitedefaultseppunct}\relax
\EndOfBibitem
\bibitem[Mueller \latin{et~al.}(2018)Mueller, Fillion-Robin, Boidol, Tian,
  Nechifor, Rampin, Corvellec, Medina, Dai, Petrushev, \latin{et~al.}
  others]{mueller2018amueller}
Mueller,~A.; Fillion-Robin,~J.-C.; Boidol,~R.; Tian,~F.; Nechifor,~P.;
  Rampin,~R.; Corvellec,~M.; Medina,~J.; Dai,~Y.; Petrushev,~B., \latin{et~al.}
   Amueller/Word\_Cloud: Wordcloud 1.5. 0. \emph{Zenodo} \textbf{2018}, \relax
\mciteBstWouldAddEndPunctfalse
\mciteSetBstMidEndSepPunct{\mcitedefaultmidpunct}
{}{\mcitedefaultseppunct}\relax
\EndOfBibitem
\bibitem[Pedregosa \latin{et~al.}(2011)Pedregosa, Varoquaux, Gramfort, Michel,
  Thirion, Grisel, Blondel, Prettenhofer, Weiss, Dubourg, \latin{et~al.}
  others]{pedregosa2011scikit}
Pedregosa,~F.; Varoquaux,~G.; Gramfort,~A.; Michel,~V.; Thirion,~B.;
  Grisel,~O.; Blondel,~M.; Prettenhofer,~P.; Weiss,~R.; Dubourg,~V.,
  \latin{et~al.}  Scikit-learn: Machine learning in Python. \emph{the Journal
  of machine Learning research} \textbf{2011}, \emph{12}, 2825--2830\relax
\mciteBstWouldAddEndPuncttrue
\mciteSetBstMidEndSepPunct{\mcitedefaultmidpunct}
{\mcitedefaultendpunct}{\mcitedefaultseppunct}\relax
\EndOfBibitem
\bibitem[Paszke \latin{et~al.}(2019)Paszke, Gross, Massa, Lerer, Bradbury,
  Chanan, Killeen, Lin, Gimelshein, Antiga, \latin{et~al.}
  others]{paszke2019pytorch}
Paszke,~A.; Gross,~S.; Massa,~F.; Lerer,~A.; Bradbury,~J.; Chanan,~G.;
  Killeen,~T.; Lin,~Z.; Gimelshein,~N.; Antiga,~L., \latin{et~al.}  Pytorch: An
  imperative style, high-performance deep learning library. \emph{Advances in
  neural information processing systems} \textbf{2019}, \emph{32}\relax
\mciteBstWouldAddEndPuncttrue
\mciteSetBstMidEndSepPunct{\mcitedefaultmidpunct}
{\mcitedefaultendpunct}{\mcitedefaultseppunct}\relax
\EndOfBibitem
\bibitem[Wang \latin{et~al.}(2019)Wang, Zheng, Ye, Gan, Li, Song, Zhou, Ma, Yu,
  Gai, \latin{et~al.} others]{wang2019deep}
Wang,~M.; Zheng,~D.; Ye,~Z.; Gan,~Q.; Li,~M.; Song,~X.; Zhou,~J.; Ma,~C.;
  Yu,~L.; Gai,~Y., \latin{et~al.}  Deep graph library: A graph-centric,
  highly-performant package for graph neural networks. \emph{arXiv preprint
  arXiv:1909.01315} \textbf{2019}, \relax
\mciteBstWouldAddEndPunctfalse
\mciteSetBstMidEndSepPunct{\mcitedefaultmidpunct}
{}{\mcitedefaultseppunct}\relax
\EndOfBibitem
\bibitem[Yao \latin{et~al.}(2019)Yao, Mao, and Luo]{yao2019graph}
Yao,~L.; Mao,~C.; Luo,~Y. Graph convolutional networks for text classification.
  Proceedings of the AAAI conference on artificial intelligence. 2019; pp
  7370--7377\relax
\mciteBstWouldAddEndPuncttrue
\mciteSetBstMidEndSepPunct{\mcitedefaultmidpunct}
{\mcitedefaultendpunct}{\mcitedefaultseppunct}\relax
\EndOfBibitem
\bibitem[Sanh \latin{et~al.}(2019)Sanh, Debut, Chaumond, and
  Wolf]{sanh2019distilbert}
Sanh,~V.; Debut,~L.; Chaumond,~J.; Wolf,~T. DistilBERT, a distilled version of
  BERT: smaller, faster, cheaper and lighter. \emph{arXiv preprint
  arXiv:1910.01108} \textbf{2019}, \relax
\mciteBstWouldAddEndPunctfalse
\mciteSetBstMidEndSepPunct{\mcitedefaultmidpunct}
{}{\mcitedefaultseppunct}\relax
\EndOfBibitem
\bibitem[Yang \latin{et~al.}(2019)Yang, Dai, Yang, Carbonell, Salakhutdinov,
  and Le]{yang2019xlnet}
Yang,~Z.; Dai,~Z.; Yang,~Y.; Carbonell,~J.; Salakhutdinov,~R.~R.; Le,~Q.~V.
  Xlnet: Generalized autoregressive pretraining for language understanding.
  \emph{Advances in neural information processing systems} \textbf{2019},
  \emph{32}\relax
\mciteBstWouldAddEndPuncttrue
\mciteSetBstMidEndSepPunct{\mcitedefaultmidpunct}
{\mcitedefaultendpunct}{\mcitedefaultseppunct}\relax
\EndOfBibitem
\bibitem[Yan \latin{et~al.}(2021)Yan, Jiang, and Dang]{yan2021named}
Yan,~R.; Jiang,~X.; Dang,~D. Named entity recognition by using
  XLNet-BiLSTM-CRF. \emph{Neural Processing Letters} \textbf{2021}, \emph{53},
  3339--3356\relax
\mciteBstWouldAddEndPuncttrue
\mciteSetBstMidEndSepPunct{\mcitedefaultmidpunct}
{\mcitedefaultendpunct}{\mcitedefaultseppunct}\relax
\EndOfBibitem
\bibitem[Raffel \latin{et~al.}(2020)Raffel, Shazeer, Roberts, Lee, Narang,
  Matena, Zhou, Li, and Liu]{raffel2020exploring}
Raffel,~C.; Shazeer,~N.; Roberts,~A.; Lee,~K.; Narang,~S.; Matena,~M.;
  Zhou,~Y.; Li,~W.; Liu,~P.~J. Exploring the limits of transfer learning with a
  unified text-to-text transformer. \emph{The Journal of Machine Learning
  Research} \textbf{2020}, \emph{21}, 5485--5551\relax
\mciteBstWouldAddEndPuncttrue
\mciteSetBstMidEndSepPunct{\mcitedefaultmidpunct}
{\mcitedefaultendpunct}{\mcitedefaultseppunct}\relax
\EndOfBibitem
\bibitem[Zhang \latin{et~al.}(2022)Zhang, Roller, Goyal, Artetxe, Chen, Chen,
  Dewan, Diab, Li, Lin, \latin{et~al.} others]{zhang2022opt}
Zhang,~S.; Roller,~S.; Goyal,~N.; Artetxe,~M.; Chen,~M.; Chen,~S.; Dewan,~C.;
  Diab,~M.; Li,~X.; Lin,~X.~V., \latin{et~al.}  Opt: Open pre-trained
  transformer language models. \emph{arXiv preprint arXiv:2205.01068}
  \textbf{2022}, \relax
\mciteBstWouldAddEndPunctfalse
\mciteSetBstMidEndSepPunct{\mcitedefaultmidpunct}
{}{\mcitedefaultseppunct}\relax
\EndOfBibitem
\bibitem[Schwartz \latin{et~al.}(2022)Schwartz, Vassilev, Greene, Perine, Burt,
  and Hall]{schwartz2022towards}
Schwartz,~R.; Vassilev,~A.; Greene,~K.; Perine,~L.; Burt,~A.; Hall,~P. Towards
  a standard for identifying and managing bias in artificial intelligence.
  \emph{NIST Special Publication} \textbf{2022}, \emph{1270}, 1--77\relax
\mciteBstWouldAddEndPuncttrue
\mciteSetBstMidEndSepPunct{\mcitedefaultmidpunct}
{\mcitedefaultendpunct}{\mcitedefaultseppunct}\relax
\EndOfBibitem
\bibitem[Wang \latin{et~al.}(2022)Wang, Kordi, Mishra, Liu, Smith, Khashabi,
  and Hajishirzi]{wang2022self}
Wang,~Y.; Kordi,~Y.; Mishra,~S.; Liu,~A.; Smith,~N.~A.; Khashabi,~D.;
  Hajishirzi,~H. Self-Instruct: Aligning Language Model with Self Generated
  Instructions. \emph{arXiv preprint arXiv:2212.10560} \textbf{2022}, \relax
\mciteBstWouldAddEndPunctfalse
\mciteSetBstMidEndSepPunct{\mcitedefaultmidpunct}
{}{\mcitedefaultseppunct}\relax
\EndOfBibitem
\bibitem[Taori \latin{et~al.}(2023)Taori, Gulrajani, Zhang, Dubois, Li,
  Guestrin, Liang, and Hashimoto]{alpaca}
Taori,~R.; Gulrajani,~I.; Zhang,~T.; Dubois,~Y.; Li,~X.; Guestrin,~C.;
  Liang,~P.; Hashimoto,~T.~B. Stanford Alpaca: An Instruction-following LLaMA
  model. \url{https://github.com/tatsu-lab/stanford_alpaca}, 2023\relax
\mciteBstWouldAddEndPuncttrue
\mciteSetBstMidEndSepPunct{\mcitedefaultmidpunct}
{\mcitedefaultendpunct}{\mcitedefaultseppunct}\relax
\EndOfBibitem
\bibitem[Choudhary and Garrity(2022)Choudhary, and
  Garrity]{choudhary2022designing}
Choudhary,~K.; Garrity,~K. Designing High-Tc Superconductors with BCS-inspired
  Screening, Density Functional Theory and Deep-learning. \emph{arXiv preprint
  arXiv:2205.00060} \textbf{2022}, \relax
\mciteBstWouldAddEndPunctfalse
\mciteSetBstMidEndSepPunct{\mcitedefaultmidpunct}
{}{\mcitedefaultseppunct}\relax
\EndOfBibitem
\bibitem[Sun \latin{et~al.}(2002)Sun, Yang, Balasubramanian, McBreen, Xia, and
  Sakai]{sun2002situ}
Sun,~X.; Yang,~X.; Balasubramanian,~M.; McBreen,~J.; Xia,~Y.; Sakai,~T. In Situ
  Investigation of Phase Transitions of Li1+ y Mn2 O 4 Spinel during Li-Ion
  Extraction and Insertion. \emph{Journal of The Electrochemical Society}
  \textbf{2002}, \emph{149}, A842\relax
\mciteBstWouldAddEndPuncttrue
\mciteSetBstMidEndSepPunct{\mcitedefaultmidpunct}
{\mcitedefaultendpunct}{\mcitedefaultseppunct}\relax
\EndOfBibitem
\bibitem[Xiao \latin{et~al.}(2021)Xiao, Beltagy, Carenini, and
  Cohan]{xiao2021primer}
Xiao,~W.; Beltagy,~I.; Carenini,~G.; Cohan,~A. Primer: Pyramid-based masked
  sentence pre-training for multi-document summarization. \emph{arXiv preprint
  arXiv:2110.08499} \textbf{2021}, \relax
\mciteBstWouldAddEndPunctfalse
\mciteSetBstMidEndSepPunct{\mcitedefaultmidpunct}
{}{\mcitedefaultseppunct}\relax
\EndOfBibitem
\bibitem[Choudhary \latin{et~al.}(2023)Choudhary, Gurunathan, DeCost, and
  Biacchi]{atomvision}
Choudhary,~K.; Gurunathan,~R.; DeCost,~B.; Biacchi,~A. AtomVision: A Machine
  Vision Library for Atomistic Images. \emph{Journal of Chemical Information
  and Modeling} \textbf{2023}, \emph{63}, 1708--1722, PMID: 36857727\relax
\mciteBstWouldAddEndPuncttrue
\mciteSetBstMidEndSepPunct{\mcitedefaultmidpunct}
{\mcitedefaultendpunct}{\mcitedefaultseppunct}\relax
\EndOfBibitem
\bibitem[Choudhary and DeCost(2021)Choudhary, and
  DeCost]{choudhary2021atomistic}
Choudhary,~K.; DeCost,~B. Atomistic line graph neural network for improved
  materials property predictions. \emph{npj Computational Materials}
  \textbf{2021}, \emph{7}, 185\relax
\mciteBstWouldAddEndPuncttrue
\mciteSetBstMidEndSepPunct{\mcitedefaultmidpunct}
{\mcitedefaultendpunct}{\mcitedefaultseppunct}\relax
\EndOfBibitem
\end{mcitethebibliography}

\end{document}